\documentclass[final,3p,times,preprint]{elsarticle}
\usepackage[utf8]{inputenc}
\usepackage{amsfonts}
\usepackage{amsmath,amssymb}
\usepackage{natbib}
\usepackage{comment}

\usepackage{bm}
\usepackage{lineno,hyperref}
\usepackage{comment}
\usepackage{xcolor}
\usepackage{booktabs}
\usepackage{multirow}%divide cells in table
\modulolinenumbers[5]
\usepackage[normalem]{ulem}

\def\nuc#1#2{\relax\ifmmode{}^{#1}{\protect\text{#2}}\else${}^{#1}$#2\fi}

\newcommand{\be}{\begin{eqnarray}}
\newcommand{\ee}{\end{eqnarray}}

\newcommand{\bwt}{\begin{widetext}}
\newcommand{\ewt}{\end{widetext}}

\begin{document}
\title{Comparison of semiclassical transfer to continuum model with Ichimura-Austern-Vincent model in medium energy knockout reactions}
%Alt: Insights into the reaction mechanisms leading to partial fusion of weakly bound nuclei

% repeat the \author .. \affiliation  etc. as needed
% \email, \thanks, \homepage, \altaffiliation all apply to the current
% author. Explanatory text should go in the []'s, actual e-mail
% address or url should go in the {}'s for \email and \homepage.
% Please use the appropriate macro foreach each type of information

% \affiliation command applies to all authors since the last
% \affiliation command. The \affiliation command should follow the
% other information
% \affiliation can be followed by \email, \homepage, \thanks as well.
\author{Jin Lei  \fnref{jinfootnote} }
\author{Angela Bonaccorso\fnref{myfootnote}}
%\tnoteref{mytitlenote}}
%\tnotetext[mytitlenote]{jin.lei@pi.infn.it, bonac@df.unipi.it}
\fntext[jinfootnote]{{ jinl@tongji.edu.cn}}
\fntext[myfootnote]{{ bonac@df.unipi.it}}

%\homepage[]{Your web page}
%\thanks{}

%\altaffiliation{Present address: Institute of Nuclear and Particle Physics, and Department of Physics and Astronomy, Ohio University, Athens, Ohio 45701, USA}
%\homepage[]{Your web page}
%\thanks{}
\address{Istituto Nazionale di Fisica Nucleare, Sezione di Pisa, Largo Pontecorvo 3, 56127 Pisa, Italy.}

%\homepage[]{Your web page}
%\thanks{}

%\address{Istituto Nazionale di Fisica Nucleare, Sezione di Pisa, Largo Pontecorvo 3, 56127 Pisa, Italy}

%Collaboration name if desired (requires use of superscriptaddress
%option in \documentclass). \noaffiliation is required (may also be
%used with the \author command).
%\collaboration can be followed by \email, \homepage, \thanks as well.
%\collaboration{}
%\noaffiliation
\begin{abstract}
 The full quantum mechanical (QM) model of inclusive breakup of  Ichimura-Austern-Vincent (IAV)  is implemented  in this paper   to calculate breakup from heavy radioactive nuclei on a $^9$Be target at intermediate energies. So far  it had been implemented and applied  only to low energy reactions with light projectiles. 
The IAV model is successful in predicting absolute cross sections among other observables. In order to get insight on the content of the model in the case of the complicated  heavy-ion reactions, results are compared  with those of the semiclassical transfer to the continuum (TC) model.  Because the TC is based on analytical formulae the dynamics of the breakup as it is contained in the rather involved IAV formalism will become more transparent. Heavy-ion reactions at high energies ($>$50A.MeV) are demanding from the computational point of view because of the high number of partial waves involved, typically around 100. The TC constitutes a useful alternative to the full QM calculations whenever predictions and/or estimates are necessary. It allows also for a systematic, fast evaluation of breakup observables. In the applications of both methods we  use state-of-the art optical potentials and structure information.  Excellent agreement is found between the calculated results of both methods and with available experimental data  which shows that the qualitative and quantitative understanding of most aspects of one nucleon breakup is well under control.
\end{abstract}

% 25.70.Mn, Projectile and target fragmentation
% 24.10.Eq 	Coupled-channel and distorted-wave models
% 25.45.-z  2H-induced reactions
% 24.87.+y 	Surrogate reactions

% insert suggested PACS numbers in braces on next line
%Pacs: 24.10.Eq, 25.70.Mn, 25.45.-z
% insert suggested keywords - APS authors don't need to do this
%\keywords{}
\date{\today}%
%\maketitle must follow title, authors, abstract, \pacs, and \keywords
\maketitle

\section{Introduction}
Breakup  in a nucleus-nucleus collision   often represents a large part of the total reaction cross section if one of the two nuclei is weakly bound. It has been studied for as long as nuclear reactions have been made, starting with the deuteron projectile case,  and its modelling can be more or less complicated depending on whether the experiment is inclusive or exclusive and on whether one considers the breakup of a nucleon or a cluster. Exclusive in our case refers to the identification of the final state of the residue by  $\gamma$-ray  detection.
From the point of view of the theoretical reaction model the challenge lies  in the description of all final state interactions between the various nucleons/nuclei present after the breakup. In principle both nuclear and Coulomb interactions should be considered but the theoretical description can be simplified  in some specific experimental conditions, for example when  a light target ion is used the Coulomb potential is often neglected. 

 To study properties of the valence nucleons in  short-lived, exotic nuclei, one-nucleon knockout at intermediate energies has been used in the last fifteen years and has largely contributed to establishing the picture of shell structure away from stability by  extracting the spectroscopic factors~\cite{Macfarlane} for the initial state wave function from the comparison of experimental data to the reaction theory predictions. 
A recent compilation of experimental knockout cross sections at intermediate energies showed a systematic trend when compared to theoretical calculations based on shell-model predictions for shell occupancy and eikonal approximation for the nucleon removal reactions~\cite{Tostevin_Gade}.  
However, this marked dependence does not seem to be supported by the results obtained with transfer reactions~\cite{Lee11,Flavigny13,Flavigny18} and quasifree
scattering with ($p$, $2p$), ($p$, $pn$), and ($e$, $e'p$) reactions \cite{PPNPall}. 
Thus there are two possibilities: i) The eikonal model is not accurate enough to describe knockout from a deeply bound state and/or ii) there other dynamical effects to be taken into account beyond those included in a peripheral model.

Here we aim at providing two alternatives to the eikonal model to help disentangling the above problems. One is fully quantum mechanical, the  so called Ichimura, Austern, and Vincent (IAV) model \cite{IAV85,AUSTERN1987125}. The IAV model has been successfully applied to study the inclusive breakup reaction induced by weakly bound nuclei, such as deuteron~\cite{Jin15} and $^{6,7}$Li~\cite{Jin18b,Jin17}, at energies around the Coulomb barrier.  The other is semiclassical, the Transfer to the Continuum  model (TC) \cite{bb,bb1,67,initst}. Semiclassical refers to the fact that the relative motion of the reaction partners is treated as a classical trajectory, which allows several simplifications in the use of coordinates and energy conservation conditions. The TC method has also been applied to the description of a large number of inclusive breakup reactions, from normal to exotic projectiles  both from deeply bound states as well as from weakly bound states \cite{67,Highensp,Tina,initst,Flomefs,geopap,enders,shane,Flavigny:2012,ppnp}. In this paper, we will compare the numerical results of these two models and compare with experimental data. Both methods  contain the correct kinematics and QM effects. The TC treats the relative motion in the semiclassical approximation but the fully QM IAV model does not and at the beginning of the paper we show  that indeed the semiclassical approximation for the relative motion is justified. On the other hand the IAV method is fully QM and in this sense it is as good as the Continuum Discretized Coupled Channel method (CDCC) which is commonly used to calculate some observables related to breakup. At the moment IAV has a larger range of applicability as it can deal  with  scattering of heavy-ions at high energy, as we argue in the following. Furthermore IAV (and TC) can both calculate the so called stripping term (NEB) while  the widely used CDCC method~\cite{AUSTERN1987125}  can only deal with the elastic breakup (diffraction). A CDCC wave function can be applied to the IAV model as described in Refs.~\cite{AUSTERN1987125,Jin19}.  However at present this is only possible at low energy and for small nuclei.

It is worth noting that the theoretical cross section depends on the description of the reaction mechanism but also on the choice of the initial state wave function. In this paper we discuss and test for the first time in the literature the validity of semi-classical approximations  with a full quantum model for the inclusive breakup reaction, thus including  both elastic and non-elastic breakup. This is very important seen the present-state-of-the-art of both experimental and theoretical knockout studies \cite{PPNPall} and our findings can open up new avenues to the understanding of nuclei with very unbalanced N/Z ratios.

We restrict ourselves to one nucleon breakup in a heavy-ion reaction. We  compare theoretical calculations to inclusive data from reactions in which the projectile-core (A$_C$=A$_P$-1)
nucleus is measured and no information is available on the target final state. Because experiments are made at relatively high energies ($>$50~A.MeV) and the core is measured intact in the forward direction,  the hypothesis is that the target  can be excited  only by the neutron-target interaction. Thus in the theoretical models the core-target scattering is considered elastic.  This is called the core-spectator model. This hypothesis is satisfied for heavy targets \cite{galin}
but it has never been proven true for light targets.

% Two theoretical frameworks will be used and compared in this paper. One is fully quantum mechanical, the  so called Ichimura, Austern, and Vincent (IAV) model \cite{IAV85,AUSTERN1987125}. The IAV model has been successfully applied to study the inclusive breakup reaction induced by weakly bound nuclei, such as deuteron~\cite{Jin15} and $^{6,7}$Li~\cite{Jin18b,Jin17}, at energies around the Coulomb barrier.  The other is semiclassical, the Transfer to the Continuum  model (TC) \cite{bb,bb1,67,initst}. Semiclassical refers to the fact that the relative motion of the reaction partners is treated as a classical trajectory, which allows several simplifications in the use of coordinates and energy conservation conditions. The TC method has also been applied to the description of a large number of inclusive breakup reactions, from normal to exotic projectiles  both from deeply bound states as well as from weakly bound states \cite{67,Highensp,Tina,initst,Flomefs,geopap,enders,shane,Flavigny:2012,ppnp}.

Because the two models have already been used in a large number of cases they are just briefly described in the supplementary material. A detailed discussion of these two models can be found in Refs~\cite{IAV85,AUSTERN1987125,Jin15,Jin15b,Jin18,Jin18b,Pot17,Carlson2016,Pot15,bb,hasan,Hasan_1979,Monaco_1985,27,Winfield85,Winfield89,Pieper78,Olmer78,PhysRevC.84.044616,PhysRevC.89.054605,BAUR1984333,bb1,67,initst,ppnp,Ravinder,firk,jer2}. Section 2 contains comparisons and discussion  of calculatated results with available experimental data and finally Sec.3 contains our conclusions and outlook.

\section{Theoretical results and comparison with experimental data}

The calculations and comparisons to data presented in this paper refer to experiments made on a $^9$Be target. Both the IAV model and the TC  use the $n-$target energy-dependent (AB) optical potential of Bonaccorso and Charity~\cite{bobme} and the single folding method of Ref.\cite{BCC1,BCC} for the core(projectile)-target potentials. The TC in general can deal with spin in both the initial  bound state and the final continuum state. The AB and DOM potentials of \cite{bobme} both contain a spin-orbit term. The IAV model can also deal with spin but so far it has been implemented numerically only for NEB. Therefore we will make comparisons between the TC and IAV results without using the spin-orbit potential. Note that also the eikonal model does neglect the spin-orbit part of  the interaction.
% , as the incident energy is very low. 
% This makes a difference

\begin{table*} [h]
 \caption {Nucleon breakup single particle cross sections in mb for the one nucleon breakup reactions $^{14}$O at 53A.MeV and $^{16}$C at 75A.MeV~ on a $^9$Be target \cite{Flavigny:2012}. Separation energies in MeV, asymptotic normalization constants C$_i$ in fm$^{-1/2}$ and cross section in mb. R$_f$ is the ratio between the experimental and IAV cross section including the shell model spectroscopic factor. Experimental and eikonal cross sections (including already the spectroscopic factors)  and spectroscopic factors from Ref.~\cite{Flavigny:2012}. See text for details.}
\centering  
%\begin{ruledtabular}
% \begin{minipage}[t]{15 cm}
%\squeezetable
\begin{tabular} {cccc|ccccccc}  
\noalign{\smallskip}
&S$_{n(p)}$&$nlj$&C$_i$&&$\sigma_{IAV}$&$\sigma_{TC}$&$C^2$S&$\sigma_{eik}$&$\sigma_{exp}$&$R_f$\\
\hline
$^{14}$O(-n)&23.12&1$p_{3/2}$&17.74&&&&3.15&&&\\ 
&&&&TOT&13.72 (6.86)&12.65&&54(0.26)&14&0.3 (0.65)\\ 
&&&&EBU&3.55&2.37&&&\\ 
&&&&NEB&10.17&10.28&&&\\ 
\hline
$^{14}$O(-p)&4.63&1$p_{1/2}$&4.20&&&&1.55&&&\\ 
&&&&TOT&33.91&30.5&&55(1.05)&58&1.10\\ 
&&&&EBU&12.50&10.3&&&&\\ 
&&&&NEB&21.41&20.2&&&&\\ 
\hline
$^{16}$C(-n)&4.25&2$s_{1/2}$&3.83&&&&0.89&&\\
&&&&TOT&58.42&47.7&&60(0.6)&36&0.7\\ 
&&&&EBU&16.09&14.3&&&&\\ 
&&&&NEB&42.33&32.4&&&&\\ 
&4.99&1$d_{5/2}$&0.90&&&&0.90&&&\\
&&&&TOT&36.29&26.9&&30(1.54)&46&1.4\\ 
&&&&EBU&10.99&7.1&&&&\\ 
&&&&NEB&25.30&19.8&&&\\ 
\hline
$^{16}$C(-p)&22.56&1$p_{3/2}$&19.26&&&&2.95&&\\
&&&&TOT&7.45&7.48&&50(0.36)&18&0.82\\ 
&&&&EBU&1.21&1.10&&&&\\ 
&&&&NEB&6.24&6.38&&&&\\ 
\noalign{\smallskip}\hline
\label{t1}
\end{tabular}
%\end{minipage}
%\end{ruledtabular}
\end{table*}

Being analytical the TC model can deal with large number of partial waves and final energies without numerical problems. Usually convergence is easily attained by taking the neutron maximum energy as twice the incident energy per nucleon, twenty n-target partial waves and the core-target impact parameter up to twice the sum of the projectile and target radii.
On the other hand for the IAV model, the numerical calculations are much more complicated compared to the TC model. Here, we use the partial wave formalism mentioned in Ref.~\cite{Jin18}, in which one needs to perform the Jacobi coordinates transformation between the incoming Jacobi coordinate, $(C+n)+T$, and outgoing Jacobi coordinate, $(n+T)+C$ (cf. Fig.1 of the supplementary material). The computing time of this kind of transformation is proportional to the cube of the maximum number of the partial waves  used in the calculations. For the current study cases on the $^9$Be target, the maximum number of partial waves has been set to be around 100 comparing to the low energy cases which only require at most 30 partial waves. In the calculations, the neutron-target energy range is chosen by stopping  the calculation when the $d\sigma/dE$ is less than $1\%$ of the maximum. Note that the IAV method has been applied so far to light projectiles and low incident energies. This is the first attempt to extend the challenging numerical calculations to higher energies and heavier projectiles.

\begin{figure}[tb]
\begin{center}
 {\centering \resizebox*{0.70\columnwidth}{!}{\includegraphics{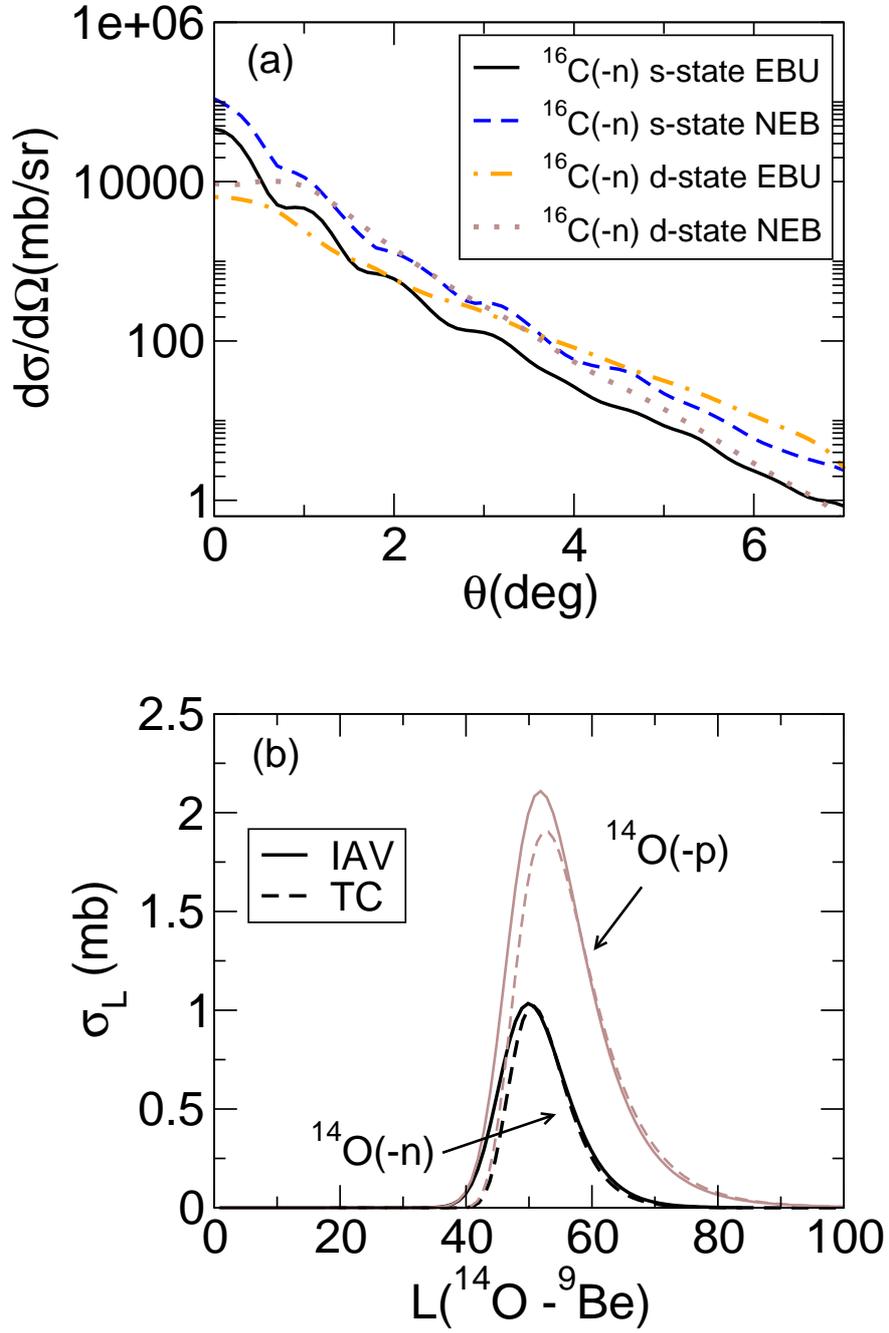}} \par}
\caption{\label{fig:dsdw}Top (a): Differential cross section angular distribution of the core in lab frame for one neutron removal reaction of $^{16}$C calculated with IAV model. Bottom (b): Projectile-target angular momentum distribution of the summed breakup reaction cross section (EBU+NEB) for the $^{14}$O + $^9$Be system, (see text for more details).}
\end{center}
\end{figure}

The results of four reactions first presented in \cite{Flavigny:2012} are discussed in this paper: both proton and neutron knockout from $^{14}$O and $^{16}$C on a $^{9}$Be target. In the case of the TC method the proton is treated by fitting its wave-function to a neutron wave function of appropriate effective separation energy as discussed in~\cite{thesis,Ravinder}. Details of separation energies, angular momentum states, asymptotic normalization constants and other important parameters are given in Table~\ref{t1}. Only the states for which experimental data and spectroscopic factors were given in Ref.~\cite{Flavigny:2012} are considered here. Namely the breakup from the valence neutron and proton states of the two projectiles and the  one-neutron removal from $^{16}$C leading to the $^{15}$C  (5/2)$^+$ bound excited state at 740 keV above its
(1/2)$^+$ ground state.  There one can already  see  that the cross sections of both EBU and NEB computed by the TC method are very close to the ones obtained by the IAV model. However to better investigate the relation between IAV and TC, the first thing we are interested in checking is the core-residual nucleus angular distribution which can  confirm whether the relative motion follows a classical path or not. This condition is necessary to make the comparison of the two methods meaningful. 
In  Fig.~\ref{fig:dsdw} (a), top part , we show the angular distribution of the core 
in the lab frame calculated by the IAV model. The solid, dashed, dotted, and dash-dotted lines are the angular distribution of $s-$state EBU, $s-$state NEB, $d-$state EBU, and $d-$state NEB, respectively. We find that the cross section decreases smoothly, it is forward peaked and basically single valued for the case of the weakly bound $d$-neutron state in $^{16}$C. The case of the $s$-state is interesting because it shows some small oscillations which remind of those seen in Ref.~\cite{bertulani04} for the breakup of the halo s-state in $^{11}$Be. According to Ref.~\cite{bertulani04} they are due to  small diffraction effects.

On the other hand in  Fig.~\ref{fig:dsdw} (b), bottom part,  the projectile-target angular momentum distribution of the breakup cross section for $^{14}$O+$^{9}$Be is presented. This figure contains the TC (dashed lines) and IAV (solid lines) results for both neutron (thick lines) and proton (thin lines) breakup. In order to make the comparison with the TC method that is based on an integral over the core-target impact parameter rather than on the sum over partial waves of the IAV model we have used the relationship $L+1/2=b_cK$ where $K$ is the momentum of relative motion at infinity. $K=8.7$ fm$^{-1}$ for the presented case. First we notice that the distributions are similar even if the neutron is strongly bound while the proton is not. The peaks are at about the same $L$ but of course the absolute values are larger for the weakly bound proton. This result suggests that in presence of such a strongly competing channel the knockout of the strongly bound particle will be suppressed in the experimental data. However the two processes are considered independent in the present existing models  and this might partially explain the large difference between the cross section experimental values and those calculated. Both models predict the same bell-shaped distribution,  because at small impact parameters breakup is suppressed by the strong core-target absorption while at large $L$ (or $b_c$) the breakup probability decreases. This kind of distribution was first predicted by Hussein-McVoy \cite{Hussein:1985} on the basis of an eikonal approximation to the IAV model. Again we see a very classical behaviour with the peak close to the strong absorption radius corresponding to a grazing collision between projectile and target. Because the wave function of the weakly bound proton has a long tail breakup extends to larger impact parameters than in the case of the strongly bound neutron. The IAV model calculates the $L$-distributions in the system of the  incoming Jacobi coordinates and  of the outgoing Jacobi coordinates, as mentioned above. The two distributions are the same but shifted by a few partial waves.  The TC considers only the core-target partial waves (impact parameters). Because of this there is a small difference with the IAV results for the low $L$. The overall agreement of the two models is excellent which demonstrates that both models describe the same physical content. In particular it shows that the use of  semiclassical approximations for heavy-ions at medium to high incident energies is justified and therefore in the future QM numerical method implementations could be simplified in as far as the relative motion wave function is concerned. It shows also that the approximations inherent  to the TC model do not change the rate of convergence.  Furthermore the absolute values of the cross sections are very close in the two models at each partial wave and in total (c.f. Table \ref{t1}).

Table~\ref{t1} contains the total cross section values and the information on the initial single particle states. The  nucleon wave functions were calculated by fitting the depth of a Woods-Saxon potential to the nucleon experimental separation energy. Radius parameter and diffuseness were 1.25~fm and 0.7~fm respectively in all but the $^{14}$O neutron case in which we had r$_0$=1.4~fm. The latter value was chosen following  Ref.~\cite{Flavigny:2012} and also  Ref.~\cite{Flavigny13} where the  $^{14}$O($d$,$p$)$^{13}$O reaction was studied. The R$_f$ value given in parentheses in the last column corresponds to the standard r$_0$=1.25~fm as explained in the following. The first column indicates the reactions studied in this paper, the separation energies and quantum numbers of the initial single particle states and the asymptotic normalization constants used in the TC calculations second to fourth column respectively. Sixth and seventh column contain the single particle total cross sections from the IAV and TC calculations as indicated. The eighth and ninth columns provide the shell model spectroscopic factors and eikonal calculation results presented in Ref.~\cite{Flavigny:2012}. In the same table, last column we show the so called "reduction factor"~\cite{PPNPall} that we indicate here as $R_f=\sigma_{exp}/(C^2S \sigma_{IAV} )$ in order to distinguish it from the strong absorption radius R$_s$. Its deviation from one is a measure of possible inaccuracies in both the reaction model and the shell model~\cite{Macfarlane}. The numbers in parenthesis next to the eikonal cross section  values are the $R_f=\sigma_{exp}/\sigma_{eik}$.  These eikonal values are given just as typical examples of results from other theoretical methods but one has to keep in mind that while we have used in this paper the same initial bound state wave functions as in Ref.~\cite{Flavigny:2012} the neutron-target and core-target potentials used for the eikonal calculations in Ref.~\cite{Flavigny:2012} were different.  Just for information of the potential readers we provide here the reduction factors obtained from the analysis of other reactions, namely R$_f$=0.5 from  $^{14}$O($p$,2$p$) \cite{61}, R$_f$=0.6 from $^{14}$O ($p$,2$p$) \cite{Mario}, R$_f$=0.73 from $^{14}$O($d$,$^3$He) \cite{Flavigny13} and R$_f$=0.54 from $^{14}$O($d$,$t$)  \cite{Flavigny13}. 

%--
\begin{figure}[tb]
\begin{center}
 {\centering \resizebox*{0.96\columnwidth}{!}{\includegraphics{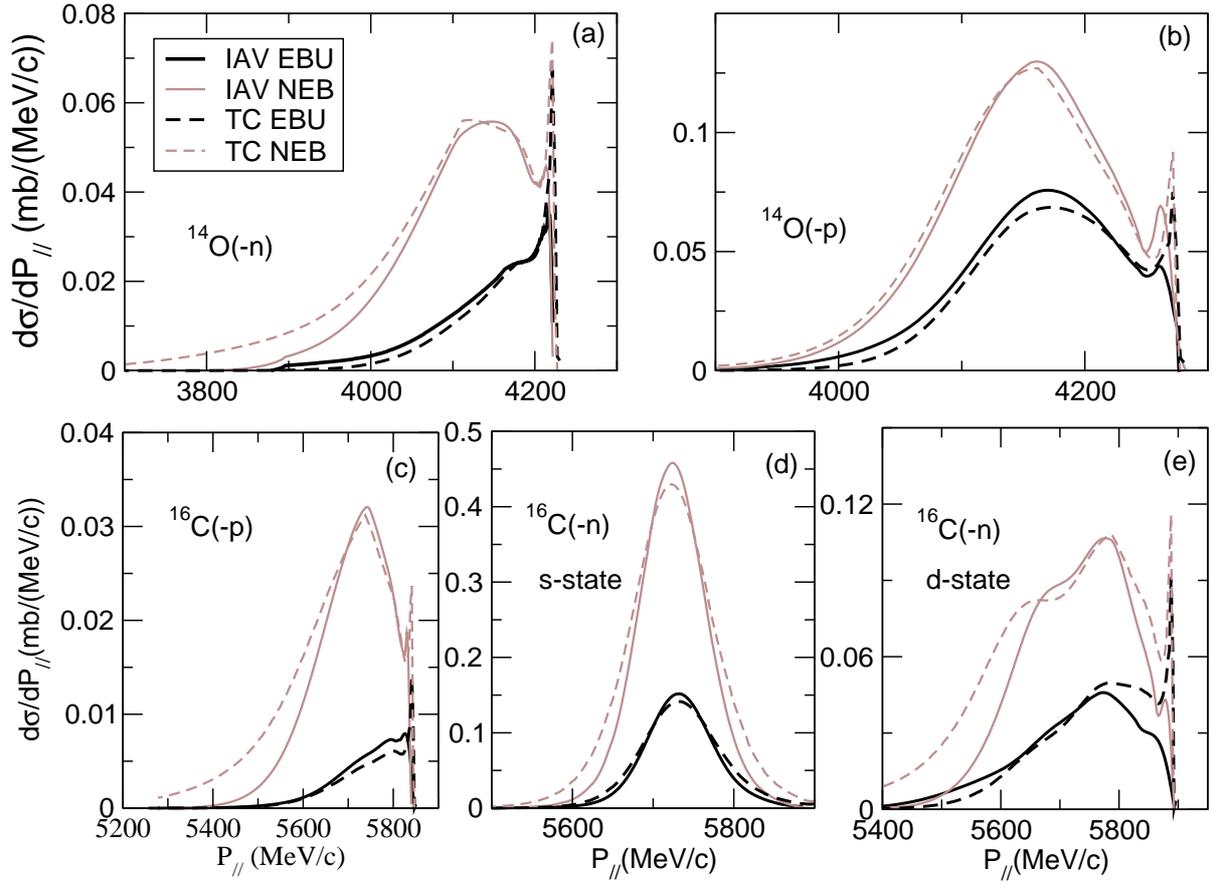}} \par}
\caption{\label{fig:comapre_IAV_TC}Core momentum distribution calculated by IAV (solid lines) and TC (dashed lines) for the reactions of (a) $^9$Be($^{14}$O,$^{13}$O)X, (b) $^9$Be($^{14}$O,$^{13}$N)X, (c) $^9$Be($^{16}$C,$^{15}$B)X, (d)
$^9$Be($^{16}$C,$^{15}$C)X where $n$ is in $s-$state, and (e) $^9$Be($^{16}$C,$^{15}$C)X where $n$ is in $d-$state. The distributions are renormalized to fit each other near the peak value.}
\end{center}
\end{figure}
%------------------------------------------

\begin{figure}[tb]
\begin{center}
 {\centering \resizebox*{0.7\columnwidth}{!}{\includegraphics{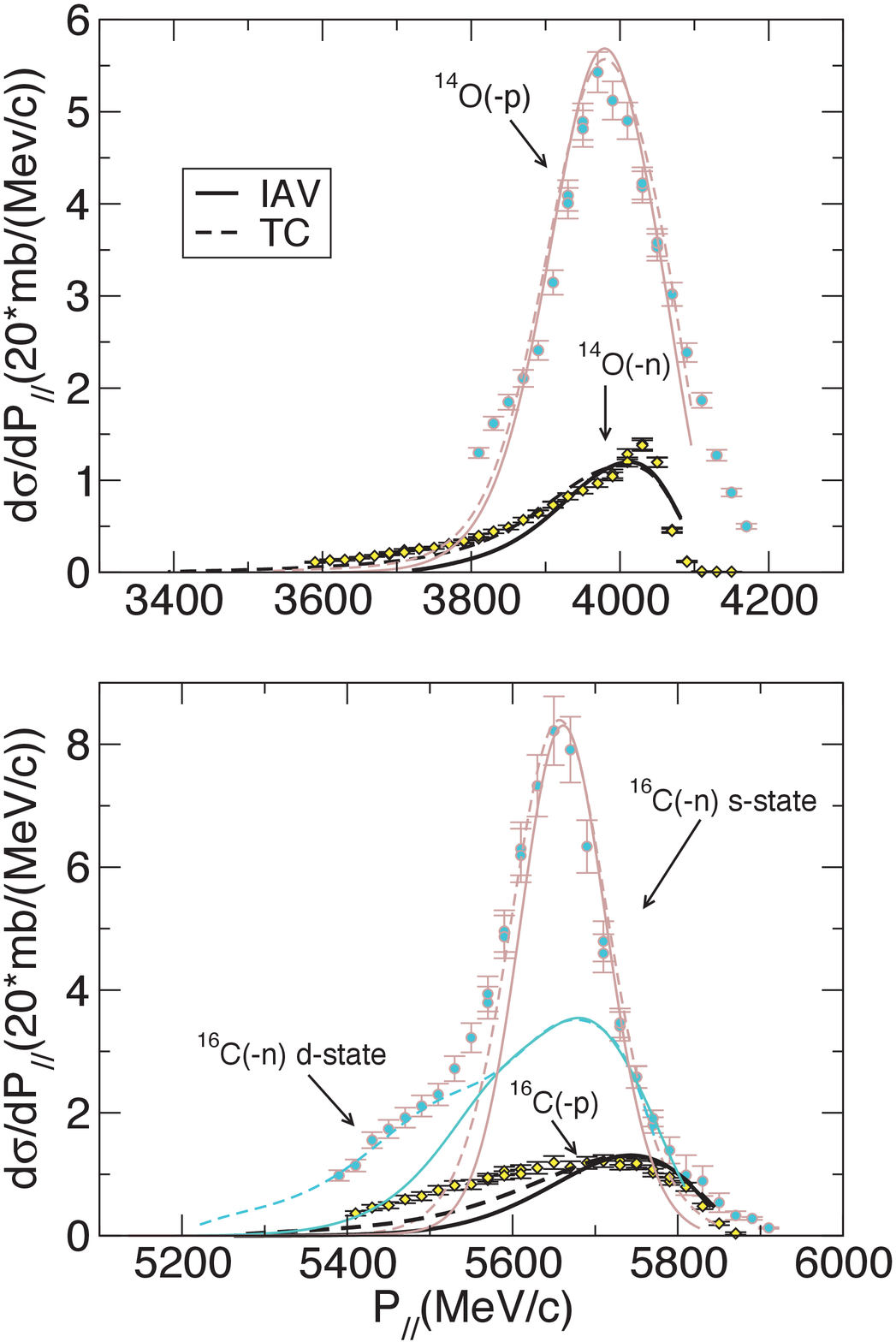}} \par}
\caption{\label{fig:comapre_data}Experimental and calculated cross section momentum distribution for the breakup reaction of (a) $^{14}$O and (b) $^{16}$C induced reactions.}
\end{center}
\end{figure}
%------------------------------------------

% ----------------------------------------
Finally in Fig.~\ref{fig:comapre_IAV_TC}  we compare the momentum distributions obtained with the two methods for both EBU and NEB and in  Fig.~\ref{fig:comapre_data} we show the total distributions sum of EBU and NEB, folded with the experimental resolution and compared to the data. The solid lines are the results obtained by IAV model and the dashed lines are calculated by TC method. We notice that the overall agreement is very good, in particular for the EBU part of the calculations the two methods give almost identical results. For the NEB the TC models provides more extended tails. These are present in the data not only those discussed in this paper, but also those in the literature  relative to exotic nuclei~\cite{thesis,ppnp} and ordinary nuclei breakup~\cite{Tina}. They appear to be due the higher n-target partial waves whose contribution is centered at higher energies in the TC results than in the IAV calculations. In Ref.~\cite{ogata} the tails were studied in detail and attributed to kinematical effects in the case of EBU from a proton target. This difference is intriguing because the total cross sections (c.f.~Table~\ref{t1}) are very close in the two methods and also the total number of partial waves necessary for convergence is basically the same in the two methods (around twelve partial waves for the present data analysis). One possible explanation is that the TC method is based on the calculation of scattering $S$-matrices for both the n-target interaction and the core-target interaction. In the IAV method, on the other hand, the EBU is calculated also via $S$-matrix formalism, while 
NEB is calculated via a source function which provides a stronger localization of the interaction in the volume region of the target.
Thus it might be possible that the IAV is more sensitive to volume aspects of the $n$-target final state interaction, while the TC is more sensitive to surface properties. 
For both methods the overall agreement with the data for the total knockout spectrum is excellent in the case of knockout of a weakly bound nucleon (-$p$ for $^{14}$O  and -$n$ for $^{16}$C), while some differences can be seen in the tails of the momentum spectra in the case of the strongly bound nucleons.

\section{Conclusions and outlook}
In this paper we have calculated and compared to data, total knockout cross sections, core parallel momentum distributions and angular momentum distributions due to elastic and non elastic breakup from
two theoretical methods: the IAV and the TC, which have been used to enlighten some semiclassical aspects of the dynamics of the knockout reaction at intermediate energies. Also the angular distributions of the cores have been calculated with the IAV method. Both the angular distributions of the core and the relative motion angular momentum distributions of the core and residual nucleus show typical semiclassical patterns. These characteristics are intrinsic of the TC model and have been confirmed by the fully quantum mechanical IAV model. Because there is also a quantitative agreement between the two methods and the data, one possible way to proceed in the future would be to implement semiclassical forms of the core-target wave functions in the IAV model. Without going to the extreme eikonal approximation suggested in Ref.~\cite{Hussein:1985}, an intermediate approach could be to use the WBK approximation for the relative motion distorted waves  which would simplify the part relative to the projectile target sum over partial waves, similar to what was done in Ref.~\cite{BAUR1984333}, but retaining the full QM evaluation of the breakup form factor. This would result in a simplification of the numerical calculations, an increased speed of them and an enlargement in the applicability of the method. In the meantime the TC has been demonstrated to be a valid and accurate alternative.

The  interactions used in this work, namely the n-$^9$Be optical potential~\cite{bobme} and the single folding model~\cite{BCC,BCC1} for the  core-target optical potential,  were tested on the n-$^{9}$Be and projectile-$^{9}$Be free particle experimental cross sections. The present results  confirm that they are accurate also in breakup calculations leading to a  good agreement with the experimental momentum distribution asymmetries and absolute cross sections. Finally with the use of standard parameters for the single particle initial state potentials, we have obtained with both methods that the extracted spectroscopic factors have at maximum a quenching of 35-40\% with respect to shell-model spectroscopic factors for the reactions studied in this paper. 

Only four reactions from Ref.~\cite{Flavigny:2012} have been analyzed but as they include neutron and proton breakup from weakly bound and strongly bound states and a smaller ($^{14}$O at 53A.MeV) and a larger ($^{16}$C at 75A.MeV) incident energy the conclusions can be considered rather general.   Encouraged by this rather satisfactory results we hope to extend our calculations to a number of other  knockout reactions such as those of Ref.~\cite{Tostevin_Gade} in order to help understanding the strong differences between experimental and theoretical cross sections discussed there.

\section*{Acknowledgments}
We are grateful to Antonio Moro and Alexandre Obertelli for a critical reading of the manuscript.
\section* {Appendix A. Supplementary material}
Supplementary material related to this article can be found on-line at
\bibliography{references}

\begin{thebibliography}{10}
\expandafter\ifx\csname url\endcsname\relax
  \def\url#1{\texttt{#1}}\fi
\expandafter\ifx\csname urlprefix\endcsname\relax\def\urlprefix{URL }\fi
\expandafter\ifx\csname href\endcsname\relax
  \def\href#1#2{#2} \def\path#1{#1}\fi

\bibitem{Macfarlane}
M.~H. Macfarlane, J.~B. French,
  \href{https://link.aps.org/doi/10.1103/RevModPhys.32.567}{Stripping reactions
  and the structure of light and intermediate nuclei}, Rev. Mod. Phys. 32
  (1960) 567--691.
\newblock \href {http://dx.doi.org/10.1103/RevModPhys.32.567}
  {\path{doi:10.1103/RevModPhys.32.567}}.
\newline\urlprefix\url{https://link.aps.org/doi/10.1103/RevModPhys.32.567}

\bibitem{Tostevin_Gade}
J.~A. Tostevin, A.~Gade,
  \href{https://link.aps.org/doi/10.1103/PhysRevC.90.057602}{Systematics of
  intermediate-energy single-nucleon removal cross sections}, Phys. Rev. C 90
  (2014) 057602.
\newblock \href {http://dx.doi.org/10.1103/PhysRevC.90.057602}
  {\path{doi:10.1103/PhysRevC.90.057602}}.
\newline\urlprefix\url{https://link.aps.org/doi/10.1103/PhysRevC.90.057602}

\bibitem{Lee11}
J.~Lee, M.~B. Tsang, D.~Bazin, D.~Coupland, V.~Henzl, D.~Henzlova, M.~Kilburn,
  W.~G. Lynch, A.~M. Rogers, A.~Sanetullaev, A.~Signoracci, Z.~Y. Sun,
  M.~Youngs, K.~Y. Chae, R.~J. Charity, H.~K. Cheung, M.~Famiano, S.~Hudan,
  P.~O'Malley, W.~A. Peters, K.~Schmitt, D.~Shapira, L.~G. Sobotka,
  \href{https://link.aps.org/doi/10.1103/PhysRevLett.104.112701}{Neutron-proton
  asymmetry dependence of spectroscopic factors in ar isotopes}, Phys. Rev.
  Lett. 104 (2010) 112701.
\newblock \href {http://dx.doi.org/10.1103/PhysRevLett.104.112701}
  {\path{doi:10.1103/PhysRevLett.104.112701}}.
\newline\urlprefix\url{https://link.aps.org/doi/10.1103/PhysRevLett.104.112701}

\bibitem{Flavigny13}
F.~Flavigny, A.~Gillibert, L.~Nalpas, A.~Obertelli, N.~Keeley, C.~Barbieri,
  D.~Beaumel, S.~Boissinot, G.~Burgunder, A.~Cipollone, A.~Corsi, J.~Gibelin,
  S.~Giron, J.~Guillot, F.~Hammache, V.~Lapoux, A.~Matta, E.~C. Pollacco,
  R.~Raabe, M.~Rejmund, N.~de~S\'ereville, A.~Shrivastava, A.~Signoracci,
  Y.~Utsuno,
  \href{https://link.aps.org/doi/10.1103/PhysRevLett.110.122503}{Limited
  asymmetry dependence of correlations from single nucleon transfer}, Phys.
  Rev. Lett. 110 (2013) 122503.
\newblock \href {http://dx.doi.org/10.1103/PhysRevLett.110.122503}
  {\path{doi:10.1103/PhysRevLett.110.122503}}.
\newline\urlprefix\url{https://link.aps.org/doi/10.1103/PhysRevLett.110.122503}

\bibitem{Flavigny18}
F.~Flavigny, N.~Keeley, A.~Gillibert, A.~Obertelli,
  \href{https://link.aps.org/doi/10.1103/PhysRevC.97.034601}{Single-particle
  strength from nucleon transfer in oxygen isotopes: Sensitivity to model
  parameters}, Phys. Rev. C 97 (2018) 034601.
\newblock \href {http://dx.doi.org/10.1103/PhysRevC.97.034601}
  {\path{doi:10.1103/PhysRevC.97.034601}}.
\newline\urlprefix\url{https://link.aps.org/doi/10.1103/PhysRevC.97.034601}

\bibitem{PPNPall}
T.~Aumann, C.~Barbieri, D.~C. Bazin, A.~Bertulani, A.~Bonaccorso, W.~H.
  Dickhoff, A.~Gade, M.~Gomez-Ramos, B.~P. Kay, A.~M. Moro, T.~Nakamura,
  A.~Obertelli, K.~Ogata, S.~Paschalis, T.~Uesaka, Quenching of single-particle
  strength from direct reactions with stable and rare-isotope beams, Progress
  in Particle and Nuclear Physics.

\bibitem{IAV85}
M.~Ichimura, N.~Austern, C.~M. Vincent,
  \href{https://link.aps.org/doi/10.1103/PhysRevC.32.431}{Equivalence of post
  and prior sum rules for inclusive breakup reactions}, Phys. Rev. C 32 (1985)
  431--439.
\newblock \href {http://dx.doi.org/10.1103/PhysRevC.32.431}
  {\path{doi:10.1103/PhysRevC.32.431}}.
\newline\urlprefix\url{https://link.aps.org/doi/10.1103/PhysRevC.32.431}

\bibitem{AUSTERN1987125}
N.~Austern, Y.~Iseri, M.~Kamimura, M.~Kawai, G.~Rawitscher, M.~Yahiro,
  \href{http://www.sciencedirect.com/science/article/pii/0370157387900949}{Continuum-discretized
  coupled-channels calculations for three-body models of deuteron-nucleus
  reactions}, Physics Reports 154~(3) (1987) 125 -- 204.
\newblock \href
  {http://dx.doi.org/https://doi.org/10.1016/0370-1573(87)90094-9}
  {\path{doi:https://doi.org/10.1016/0370-1573(87)90094-9}}.
\newline\urlprefix\url{http://www.sciencedirect.com/science/article/pii/0370157387900949}

\bibitem{Jin15}
J.~Lei, A.~M. Moro,
  \href{https://link.aps.org/doi/10.1103/PhysRevC.92.044616}{Reexamining
  closed-form formulae for inclusive breakup: Application to deuteron- and
  $^{6}\mathrm{Li}$-induced reactions}, Phys. Rev. C 92 (2015) 044616.
\newblock \href {http://dx.doi.org/10.1103/PhysRevC.92.044616}
  {\path{doi:10.1103/PhysRevC.92.044616}}.
\newline\urlprefix\url{https://link.aps.org/doi/10.1103/PhysRevC.92.044616}

\bibitem{Jin18b}
J.~Lei, A.~M. Moro,
  \href{https://link.aps.org/doi/10.1103/PhysRevC.97.011601}{Post-prior
  equivalence for transfer reactions with complex potentials}, Phys. Rev. C 97
  (2018) 011601.
\newblock \href {http://dx.doi.org/10.1103/PhysRevC.97.011601}
  {\path{doi:10.1103/PhysRevC.97.011601}}.
\newline\urlprefix\url{https://link.aps.org/doi/10.1103/PhysRevC.97.011601}

\bibitem{Jin17}
J.~Lei, A.~M. Moro,
  \href{https://link.aps.org/doi/10.1103/PhysRevC.95.044605}{Comprehensive
  analysis of large $\ensuremath{\alpha}$ yields observed in
  $^{6}\mathbf{Li}$-induced reactions}, Phys. Rev. C 95 (2017) 044605.
\newblock \href {http://dx.doi.org/10.1103/PhysRevC.95.044605}
  {\path{doi:10.1103/PhysRevC.95.044605}}.
\newline\urlprefix\url{https://link.aps.org/doi/10.1103/PhysRevC.95.044605}

\bibitem{bb}
A.~Bonaccorso, D.~M. Brink,
  \href{https://link.aps.org/doi/10.1103/PhysRevC.38.1776}{Nucleon transfer to
  continuum states}, Phys. Rev. C 38 (1988) 1776--1786.
\newblock \href {http://dx.doi.org/10.1103/PhysRevC.38.1776}
  {\path{doi:10.1103/PhysRevC.38.1776}}.
\newline\urlprefix\url{https://link.aps.org/doi/10.1103/PhysRevC.38.1776}

\bibitem{bb1}
A.~Bonaccorso, D.~M. Brink,
  \href{https://link.aps.org/doi/10.1103/PhysRevC.43.299}{Absorption versus
  breakup in heavy-ion reactions}, Phys. Rev. C 43 (1991) 299--310.
\newblock \href {http://dx.doi.org/10.1103/PhysRevC.43.299}
  {\path{doi:10.1103/PhysRevC.43.299}}.
\newline\urlprefix\url{https://link.aps.org/doi/10.1103/PhysRevC.43.299}

\bibitem{67}
A.~Bonaccorso, D.~M. Brink,
  \href{https://link.aps.org/doi/10.1103/PhysRevC.44.1559}{Stripping to the
  continuum of $^{208}\mathrm{Pb}$}, Phys. Rev. C 44 (1991) 1559--1568.
\newblock \href {http://dx.doi.org/10.1103/PhysRevC.44.1559}
  {\path{doi:10.1103/PhysRevC.44.1559}}.
\newline\urlprefix\url{https://link.aps.org/doi/10.1103/PhysRevC.44.1559}

\bibitem{initst}
A.~Bonaccorso,
  \href{https://link.aps.org/doi/10.1103/PhysRevC.60.054604}{Initial state
  dependence of the breakup of weakly bound carbon isotopes}, Phys. Rev. C 60
  (1999) 054604.
\newblock \href {http://dx.doi.org/10.1103/PhysRevC.60.054604}
  {\path{doi:10.1103/PhysRevC.60.054604}}.
\newline\urlprefix\url{https://link.aps.org/doi/10.1103/PhysRevC.60.054604}

\bibitem{Highensp}
A.~Bonaccorso, \href{https://link.aps.org/doi/10.1103/PhysRevC.51.822}{High
  energy single particle states in the continuum}, Phys. Rev. C 51 (1995)
  822--835.
\newblock \href {http://dx.doi.org/10.1103/PhysRevC.51.822}
  {\path{doi:10.1103/PhysRevC.51.822}}.
\newline\urlprefix\url{https://link.aps.org/doi/10.1103/PhysRevC.51.822}

\bibitem{Tina}
A.~Bonaccorso, I.~Lhenry, T.~Suomij\"arvi,
  \href{https://link.aps.org/doi/10.1103/PhysRevC.49.329}{Inclusive spectra of
  stripping reactions induced by heavy ions}, Phys. Rev. C 49 (1994) 329--337.
\newblock \href {http://dx.doi.org/10.1103/PhysRevC.49.329}
  {\path{doi:10.1103/PhysRevC.49.329}}.
\newline\urlprefix\url{https://link.aps.org/doi/10.1103/PhysRevC.49.329}

\bibitem{Flomefs}
A.~Bonaccorso, F.~Carstoiu,
  \href{https://link.aps.org/doi/10.1103/PhysRevC.61.034605}{Final state
  interaction effects in breakup reactions of halo nuclei}, Phys. Rev. C 61
  (2000) 034605.
\newblock \href {http://dx.doi.org/10.1103/PhysRevC.61.034605}
  {\path{doi:10.1103/PhysRevC.61.034605}}.
\newline\urlprefix\url{https://link.aps.org/doi/10.1103/PhysRevC.61.034605}

\bibitem{geopap}
A.~Bonaccorso, G.~F. Bertsch,
  \href{https://link.aps.org/doi/10.1103/PhysRevC.63.044604}{Comparison of
  transfer-to-continuum and eikonal models of projectile fragmentation
  reactions}, Phys. Rev. C 63 (2001) 044604.
\newblock \href {http://dx.doi.org/10.1103/PhysRevC.63.044604}
  {\path{doi:10.1103/PhysRevC.63.044604}}.
\newline\urlprefix\url{https://link.aps.org/doi/10.1103/PhysRevC.63.044604}

\bibitem{enders}
J.~Enders, A.~Bauer, D.~Bazin, A.~Bonaccorso, B.~A. Brown, T.~Glasmacher, P.~G.
  Hansen, V.~Maddalena, K.~L. Miller, A.~Navin, B.~M. Sherrill, J.~A. Tostevin,
  \href{https://link.aps.org/doi/10.1103/PhysRevC.65.034318}{Single-neutron
  knockout from ${}^{34,35}\mathrm{Si}$ and ${}^{37}\mathrm{S}$}, Phys. Rev. C
  65 (2002) 034318.
\newblock \href {http://dx.doi.org/10.1103/PhysRevC.65.034318}
  {\path{doi:10.1103/PhysRevC.65.034318}}.
\newline\urlprefix\url{https://link.aps.org/doi/10.1103/PhysRevC.65.034318}

\bibitem{shane}
R.~Shane, R.~J. Charity, L.~G. Sobotka, D.~Bazin, B.~A. Brown, A.~Gade, G.~F.
  Grinyer, S.~McDaniel, A.~Ratkiewicz, D.~Weisshaar, A.~Bonaccorso, J.~A.
  Tostevin, \href{https://link.aps.org/doi/10.1103/PhysRevC.85.064612}{Proton
  and neutron knockout from ${}^{36}$ca}, Phys. Rev. C 85 (2012) 064612.
\newblock \href {http://dx.doi.org/10.1103/PhysRevC.85.064612}
  {\path{doi:10.1103/PhysRevC.85.064612}}.
\newline\urlprefix\url{https://link.aps.org/doi/10.1103/PhysRevC.85.064612}

\bibitem{Flavigny:2012}
F.~Flavigny, A.~Obertelli, A.~Bonaccorso, G.~F. Grinyer, C.~Louchart,
  L.~Nalpas, A.~Signoracci, Nonsudden limits of heavy-ion induced knockout
  reactions, Phys. Rev. Lett. 108 (2012) 252501.
\newblock \href {http://dx.doi.org/10.1103/PhysRevLett.108.252501}
  {\path{doi:10.1103/PhysRevLett.108.252501}}.

\bibitem{ppnp}
A.~Bonaccorso,
  \href{http://www.sciencedirect.com/science/article/pii/S014664101830005X}{Direct
  reaction theories for exotic nuclei: An introduction via semi-classical
  methods}, Progress in Particle and Nuclear Physics 101 (2018) 1 -- 54.
\newblock \href {http://dx.doi.org/https://doi.org/10.1016/j.ppnp.2018.01.005}
  {\path{doi:https://doi.org/10.1016/j.ppnp.2018.01.005}}.
\newline\urlprefix\url{http://www.sciencedirect.com/science/article/pii/S014664101830005X}

\bibitem{Jin19}
J.~Lei, A.~M. Moro,
  \href{https://link.aps.org/doi/10.1103/PhysRevLett.123.232501}{Unraveling the
  reaction mechanisms leading to partial fusion of weakly bound nuclei}, Phys.
  Rev. Lett. 123 (2019) 232501.
\newblock \href {http://dx.doi.org/10.1103/PhysRevLett.123.232501}
  {\path{doi:10.1103/PhysRevLett.123.232501}}.
\newline\urlprefix\url{https://link.aps.org/doi/10.1103/PhysRevLett.123.232501}

\bibitem{galin}
Y.~Perier, B.~Lott, J.~Galin, E.~Linard, M.~Morjean, N.~Orr, A.~Poghaire,
  B.~Quednau, A.~Villari,
  \href{http://www.sciencedirect.com/science/article/pii/S0370269399006796}{Interplay
  between the neutron halo structure and reaction mechanisms in collisions of
  35 mev/nucleon 6he with au}, Physics Letters B 459~(1) (1999) 55 -- 60.
\newblock \href
  {http://dx.doi.org/https://doi.org/10.1016/S0370-2693(99)00679-6}
  {\path{doi:https://doi.org/10.1016/S0370-2693(99)00679-6}}.
\newline\urlprefix\url{http://www.sciencedirect.com/science/article/pii/S0370269399006796}

\bibitem{Jin15b}
J.~Lei, A.~M. Moro,
  \href{https://link.aps.org/doi/10.1103/PhysRevC.92.061602}{Numerical
  assessment of post-prior equivalence for inclusive breakup reactions}, Phys.
  Rev. C 92 (2015) 061602.
\newblock \href {http://dx.doi.org/10.1103/PhysRevC.92.061602}
  {\path{doi:10.1103/PhysRevC.92.061602}}.
\newline\urlprefix\url{https://link.aps.org/doi/10.1103/PhysRevC.92.061602}

\bibitem{Jin18}
J.~Lei, \href{https://link.aps.org/doi/10.1103/PhysRevC.97.034628}{Inclusive
  breakup calculations in angular momentum basis: Application to
  $^{7}\mathrm{Li}+^{58}\mathrm{Ni}$}, Phys. Rev. C 97 (2018) 034628.
\newblock \href {http://dx.doi.org/10.1103/PhysRevC.97.034628}
  {\path{doi:10.1103/PhysRevC.97.034628}}.
\newline\urlprefix\url{https://link.aps.org/doi/10.1103/PhysRevC.97.034628}

\bibitem{Pot17}
G.~Potel, G.~Perdikakis, B.~V. Carlson, M.~C. Atkinson, W.~H. Dickhoff, J.~E.
  Escher, M.~S. Hussein, J.~Lei, W.~Li, A.~O. Macchiavelli, A.~M. Moro, F.~M.
  Nunes, S.~D. Pain, J.~Rotureau,
  \href{https://doi.org/10.1140/epja/i2017-12371-9}{Toward a complete theory
  for predicting inclusive deuteron breakup away from stability}, Eur. Phys. J.
  A 53~(9) (2017) 178.
\newblock \href {http://dx.doi.org/10.1140/epja/i2017-12371-9}
  {\path{doi:10.1140/epja/i2017-12371-9}}.
\newline\urlprefix\url{https://doi.org/10.1140/epja/i2017-12371-9}

\bibitem{Carlson2016}
B.~V. Carlson, R.~Capote, M.~Sin,
  \href{http://dx.doi.org/10.1007/s00601-016-1054-8}{Inclusive proton emission
  spectra from deuteron breakup reactions}, Few-Body Syst. 57~(5) (2016)
  307--314.
\newblock \href {http://dx.doi.org/10.1007/s00601-016-1054-8}
  {\path{doi:10.1007/s00601-016-1054-8}}.
\newline\urlprefix\url{http://dx.doi.org/10.1007/s00601-016-1054-8}

\bibitem{Pot15}
G.~Potel, F.~M. Nunes, I.~J. Thompson,
  \href{http://0-link.aps.org.fama.us.es/doi/10.1103/PhysRevC.92.034611}{Establishing
  a theory for deuteron-induced surrogate reactions}, Phys. Rev. C 92 (2015)
  034611.
\newblock \href {http://dx.doi.org/10.1103/PhysRevC.92.034611}
  {\path{doi:10.1103/PhysRevC.92.034611}}.
\newline\urlprefix\url{http://0-link.aps.org.fama.us.es/doi/10.1103/PhysRevC.92.034611}

\bibitem{hasan}
H.~Hasan,
  \href{https://ora.ox.ac.uk/?utf8=%E2%9C%93&q=Hasan%2C+Hashima&search_field=author}{Study
  of reactions between heavy nuclei}, Ph.D. thesis, University of Oxford
  (1976).
\newline\urlprefix\url{https://ora.ox.ac.uk/?utf8=%E2%9C%93&q=Hasan%2C+Hashima&search_field=author}

\bibitem{Hasan_1979}
H.~Hasan, D.~M. Brink,
  \href{https://doi.org/10.1088%2F0305-4616%2F5%2F6%2F005}{The transfer
  amplitude and angular distributions in heavy-ion reactions}, Journal of
  Physics G: Nuclear Physics 5~(6) (1979) 771--779.
\newblock \href {http://dx.doi.org/10.1088/0305-4616/5/6/005}
  {\path{doi:10.1088/0305-4616/5/6/005}}.
\newline\urlprefix\url{https://doi.org/10.1088%2F0305-4616%2F5%2F6%2F005}

\bibitem{Monaco_1985}
L.~{Lo Monaco}, D.~M. Brink,
  \href{https://doi.org/10.1088%2F0305-4616%2F11%2F8%2F010}{Perturbation
  approach to nucleon transfer in heavy-ion reactions}, Journal of Physics G:
  Nuclear Physics 11~(8) (1985) 935--952.
\newblock \href {http://dx.doi.org/10.1088/0305-4616/11/8/010}
  {\path{doi:10.1088/0305-4616/11/8/010}}.
\newline\urlprefix\url{https://doi.org/10.1088%2F0305-4616%2F11%2F8%2F010}

\bibitem{27}
A.~Bonaccorso, D.~M. Brink, L.~L. Monaco,
  \href{https://doi.org/10.1088%2F0305-4616%2F13%2F11%2F013}{Nucleon transfer
  in heavy-ion reactions: energy dependence of the cross section}, Journal of
  Physics G: Nuclear Physics 13~(11) (1987) 1407--1428.
\newblock \href {http://dx.doi.org/10.1088/0305-4616/13/11/013}
  {\path{doi:10.1088/0305-4616/13/11/013}}.
\newline\urlprefix\url{https://doi.org/10.1088%2F0305-4616%2F13%2F11%2F013}

\bibitem{Winfield85}
J.~Winfield, N.~Jelley, W.~Rae, C.~Woods,
  \href{http://www.sciencedirect.com/science/article/pii/0375947485902271}{Spectroscopic-factor
  discrepancies in (9be, 10b) for different ejectile excitations}, Nuclear
  Physics A 437~(1) (1985) 65 -- 92.
\newblock \href
  {http://dx.doi.org/https://doi.org/10.1016/0375-9474(85)90227-1}
  {\path{doi:https://doi.org/10.1016/0375-9474(85)90227-1}}.
\newline\urlprefix\url{http://www.sciencedirect.com/science/article/pii/0375947485902271}

\bibitem{Winfield89}
J.~S. Winfield, E.~Adamides, S.~M. Austin, G.~M. Crawley, M.~F. Mohar, C.~A.
  Ogilvie, B.~Sherrill, M.~Torres, G.~Yoo, A.~Nadasen,
  \href{https://link.aps.org/doi/10.1103/PhysRevC.39.1395}{$^{12}\mathrm{induced}$
  single particle transfer reactions at e/a=50 mev}, Phys. Rev. C 39 (1989)
  1395--1401.
\newblock \href {http://dx.doi.org/10.1103/PhysRevC.39.1395}
  {\path{doi:10.1103/PhysRevC.39.1395}}.
\newline\urlprefix\url{https://link.aps.org/doi/10.1103/PhysRevC.39.1395}

\bibitem{Pieper78}
S.~C. Pieper, M.~H. Macfarlane, D.~H. Gloeckner, D.~G. Kovar, F.~D. Becchetti,
  B.~G. Harvey, D.~L. Hendrie, H.~Homeyer, J.~Mahoney, F.~P\"uhlhofer, W.~von
  Oertzen, M.~S. Zisman,
  \href{http://link.aps.org/doi/10.1103/PhysRevC.18.180}{Energy dependence of
  elastic scattering and one-nucleon transfer reactions induced by
  $^{16}\mathrm{O}$ on $^{208}\mathrm{Pb}$. i}, Phys. Rev. C 18 (1978)
  180--204.
\newblock \href {http://dx.doi.org/10.1103/PhysRevC.18.180}
  {\path{doi:10.1103/PhysRevC.18.180}}.
\newline\urlprefix\url{http://link.aps.org/doi/10.1103/PhysRevC.18.180}

\bibitem{Olmer78}
C.~Olmer, M.~Mermaz, M.~Buenerd, C.~K. Gelbke, D.~L. Hendrie, J.~Mahoney, D.~K.
  Scott, M.~H. Macfarlane, S.~C. Pieper,
  \href{http://link.aps.org/doi/10.1103/PhysRevC.18.205}{Energy dependence of
  elastic scattering and one-nucleon transfer reactions induced by
  $^{16}\mathrm{O}$ on $^{208}\mathrm{Pb}$. ii}, Phys. Rev. C 18 (1978)
  205--222.
\newblock \href {http://dx.doi.org/10.1103/PhysRevC.18.205}
  {\path{doi:10.1103/PhysRevC.18.205}}.
\newline\urlprefix\url{http://link.aps.org/doi/10.1103/PhysRevC.18.205}

\bibitem{PhysRevC.84.044616}
A.~M. Mukhamedzhanov,
  \href{https://link.aps.org/doi/10.1103/PhysRevC.84.044616}{Theory of deuteron
  stripping: From surface integrals to a generalized $r$-matrix approach},
  Phys. Rev. C 84 (2011) 044616.
\newblock \href {http://dx.doi.org/10.1103/PhysRevC.84.044616}
  {\path{doi:10.1103/PhysRevC.84.044616}}.
\newline\urlprefix\url{https://link.aps.org/doi/10.1103/PhysRevC.84.044616}

\bibitem{PhysRevC.89.054605}
J.~E. Escher, I.~J. Thompson, G.~Arbanas, C.~Elster, V.~Eremenko, L.~Hlophe,
  F.~M. Nunes,
  \href{https://link.aps.org/doi/10.1103/PhysRevC.89.054605}{Reexamining
  surface-integral formulations for one-nucleon transfers to bound and
  resonance states}, Phys. Rev. C 89 (2014) 054605.
\newblock \href {http://dx.doi.org/10.1103/PhysRevC.89.054605}
  {\path{doi:10.1103/PhysRevC.89.054605}}.
\newline\urlprefix\url{https://link.aps.org/doi/10.1103/PhysRevC.89.054605}

\bibitem{BAUR1984333}
G.~Baur, F.~Rösel, D.~Trautmann, R.~Shyam,
  \href{http://www.sciencedirect.com/science/article/pii/0370157384901388}{Fragmentation
  processes in nuclear reactions}, Physics Reports 111~(5) (1984) 333 -- 371.
\newblock \href
  {http://dx.doi.org/https://doi.org/10.1016/0370-1573(84)90138-8}
  {\path{doi:https://doi.org/10.1016/0370-1573(84)90138-8}}.
\newline\urlprefix\url{http://www.sciencedirect.com/science/article/pii/0370157384901388}

\bibitem{Ravinder}
R.~Kumar, A.~Bonaccorso, Dynamical effects in proton breakup from exotic
  nuclei, Phys. Rev. C 84 (2011) 014613.
\newblock \href {http://dx.doi.org/10.1103/PhysRevC.84.014613}
  {\path{doi:10.1103/PhysRevC.84.014613}}.

\bibitem{firk}
F.~W.~K. Firk, Introduction to relativistic collisions\href
  {http://dx.doi.org/https://arxiv.org/abs/1011.1943}
  {\path{doi:https://arxiv.org/abs/1011.1943}}.

\bibitem{jer2}
J.~Margueron, A.~Bonaccorso, D.~Brink, A non-perturbative approach to halo
  breakup, Nuclear Physics A 720~(3) (2003) 337 -- 353.
\newblock \href
  {http://dx.doi.org/https://doi.org/10.1016/S0375-9474(03)01092-3}
  {\path{doi:https://doi.org/10.1016/S0375-9474(03)01092-3}}.

\bibitem{bobme}
A.~Bonaccorso, R.~J. Charity,
  \href{https://link.aps.org/doi/10.1103/PhysRevC.89.024619}{Optical potential
  for the n-${}^{9}$be reaction}, Phys. Rev. C 89 (2014) 024619.
\newblock \href {http://dx.doi.org/10.1103/PhysRevC.89.024619}
  {\path{doi:10.1103/PhysRevC.89.024619}}.
\newline\urlprefix\url{https://link.aps.org/doi/10.1103/PhysRevC.89.024619}

\bibitem{BCC1}
A.~Bonaccorso, F.~Carstoiu, R.~J. Charity,
  \href{https://link.aps.org/doi/10.1103/PhysRevC.94.034604}{Imaginary part of
  the $^{9}\mathrm{C}\text{\ensuremath{-}}^{9}\mathrm{Be}$ single-folded
  optical potential}, Phys. Rev. C 94 (2016) 034604.
\newblock \href {http://dx.doi.org/10.1103/PhysRevC.94.034604}
  {\path{doi:10.1103/PhysRevC.94.034604}}.
\newline\urlprefix\url{https://link.aps.org/doi/10.1103/PhysRevC.94.034604}

\bibitem{BCC}
A.~Bonaccorso, F.~Carstoiu, R.~Charity, R.~Kumar, G.~Salvioni, {Differences
  Between a Single- and a Double-Folding Nucleus-$^\mathbf{9}$ Be Optical
  Potential}, Few Body Syst. 57~(5) (2016) 331--336.
\newblock \href {http://dx.doi.org/10.1007/s00601-016-1082-4}
  {\path{doi:10.1007/s00601-016-1082-4}}.

\bibitem{thesis}
G.~Salvioni, A systematic study of knockout reactions from exotic nuclei with
  anomalous ratio n/z, Master's thesis, University of Pisa (2014).
\newblock \href
  {http://dx.doi.org/https://etd.adm.unipi.it/theses/available/etd-06242014-111952/}
  {\path{doi:https://etd.adm.unipi.it/theses/available/etd-06242014-111952/}}.

\bibitem{bertulani04}
C.~A. Bertulani, P.~G. Hansen,
  \href{https://link.aps.org/doi/10.1103/PhysRevC.70.034609}{Momentum
  distributions in stripping reactions of radioactive projectiles at
  intermediate energies}, Phys. Rev. C 70 (2004) 034609.
\newblock \href {http://dx.doi.org/10.1103/PhysRevC.70.034609}
  {\path{doi:10.1103/PhysRevC.70.034609}}.
\newline\urlprefix\url{https://link.aps.org/doi/10.1103/PhysRevC.70.034609}

\bibitem{Hussein:1985}
M.~Hussein, K.~McVoy, Inclusive projectile fragmentation in the spectator
  model, Nucl. Phys. A 445 (1985) 124 -- 139.
\newblock \href
  {http://dx.doi.org/https://doi.org/10.1016/0375-9474(85)90364-1}
  {\path{doi:https://doi.org/10.1016/0375-9474(85)90364-1}}.

\bibitem{61}
S.~Kawase, T.~Uesaka, T.~L. Tang, D.~Beaumel, M.~Dozono, T.~Fukunaga, T.~Fujii,
  N.~Fukuda, A.~Galindo-Uribarri, S.~Hwang, N.~Inabe, T.~Kawabata, T.~Kawahara,
  W.~Kim, K.~Kisamori, M.~Kobayashi, T.~Kubo, Y.~Kubota, K.~Kusaka, C.~Lee,
  Y.~Maeda, H.~Matsubara, S.~Michimasa, H.~Miya, T.~Noro, Y.~Nozawa,
  A.~Obertelli, K.~Ogata, S.~Ota, E.~Padilla-Rodal, S.~Sakaguchi, H.~Sakai,
  M.~Sasano, S.~Shimoura, S.~Stepanyan, H.~Suzuki, T.~Suzuki, M.~Takaki,
  H.~Takeda, A.~Tamii, H.~Tokieda, T.~Wakasa, T.~Wakui, K.~Yako, J.~Yasuda,
  Y.~Yanagisawa, R.~Yokoyama, K.~Yoshida, K.~Yoshida, J.~Zenihiro,
  \href{https://doi.org/10.1093/ptep/pty011}{{Exclusive quasi-free proton
  knockout from oxygen isotopes at intermediate energies}}, Progress of
  Theoretical and Experimental Physics 2018~(2), 021D01.
\newblock \href
  {http://arxiv.org/abs/https://academic.oup.com/ptep/article-pdf/2018/2/021D01/24195568/pty011.pdf}
  {\path{arXiv:https://academic.oup.com/ptep/article-pdf/2018/2/021D01/24195568/pty011.pdf}},
  \href {http://dx.doi.org/10.1093/ptep/pty011}
  {\path{doi:10.1093/ptep/pty011}}.
\newline\urlprefix\url{https://doi.org/10.1093/ptep/pty011}

\bibitem{Mario}
M.~Gomez-Ramos, A.~Moro,
  \href{http://www.sciencedirect.com/science/article/pii/S0370269318306749}{Binding-energy
  independence of reduced spectroscopic strengths derived from (p,2p) and
  (p,pn) reactions with nitrogen and oxygen isotopes}, Physics Letters B 785
  (2018) 511 -- 516.
\newblock \href
  {http://dx.doi.org/https://doi.org/10.1016/j.physletb.2018.08.058}
  {\path{doi:https://doi.org/10.1016/j.physletb.2018.08.058}}.
\newline\urlprefix\url{http://www.sciencedirect.com/science/article/pii/S0370269318306749}

\bibitem{ogata}
K.~Ogata, K.~Yoshida, K.~Minomo,
  \href{https://link.aps.org/doi/10.1103/PhysRevC.92.034616}{Asymmetry of the
  parallel momentum distribution of ($p,pn$) reaction residues}, Phys. Rev. C
  92 (2015) 034616.
\newblock \href {http://dx.doi.org/10.1103/PhysRevC.92.034616}
  {\path{doi:10.1103/PhysRevC.92.034616}}.
\newline\urlprefix\url{https://link.aps.org/doi/10.1103/PhysRevC.92.034616}

\end{thebibliography}


\begin{thebibliography}{10}
\expandafter\ifx\csname url\endcsname\relax
  \def\url#1{\texttt{#1}}\fi
\expandafter\ifx\csname urlprefix\endcsname\relax\def\urlprefix{URL }\fi
\expandafter\ifx\csname href\endcsname\relax
  \def\href#1#2{#2} \def\path#1{#1}\fi

\bibitem{IAV85}
M.~Ichimura, N.~Austern, C.~M. Vincent,
  \href{https://link.aps.org/doi/10.1103/PhysRevC.32.431}{Equivalence of post
  and prior sum rules for inclusive breakup reactions}, Phys. Rev. C 32 (1985)
  431--439.
\newblock \href {http://dx.doi.org/10.1103/PhysRevC.32.431}
  {\path{doi:10.1103/PhysRevC.32.431}}.
\newline\urlprefix\url{https://link.aps.org/doi/10.1103/PhysRevC.32.431}

\bibitem{AUSTERN1987125}
N.~Austern, Y.~Iseri, M.~Kamimura, M.~Kawai, G.~Rawitscher, M.~Yahiro,
  \href{http://www.sciencedirect.com/science/article/pii/0370157387900949}{Continuum-discretized
  coupled-channels calculations for three-body models of deuteron-nucleus
  reactions}, Physics Reports 154~(3) (1987) 125 -- 204.
\newblock \href
  {http://dx.doi.org/https://doi.org/10.1016/0370-1573(87)90094-9}
  {\path{doi:https://doi.org/10.1016/0370-1573(87)90094-9}}.
\newline\urlprefix\url{http://www.sciencedirect.com/science/article/pii/0370157387900949}

\bibitem{Jin15}
J.~Lei, A.~M. Moro,
  \href{https://link.aps.org/doi/10.1103/PhysRevC.92.044616}{Reexamining
  closed-form formulae for inclusive breakup: Application to deuteron- and
  $^{6}\mathrm{Li}$-induced reactions}, Phys. Rev. C 92 (2015) 044616.
\newblock \href {http://dx.doi.org/10.1103/PhysRevC.92.044616}
  {\path{doi:10.1103/PhysRevC.92.044616}}.
\newline\urlprefix\url{https://link.aps.org/doi/10.1103/PhysRevC.92.044616}

\bibitem{Jin15b}
J.~Lei, A.~M. Moro,
  \href{https://link.aps.org/doi/10.1103/PhysRevC.92.061602}{Numerical
  assessment of post-prior equivalence for inclusive breakup reactions}, Phys.
  Rev. C 92 (2015) 061602.
\newblock \href {http://dx.doi.org/10.1103/PhysRevC.92.061602}
  {\path{doi:10.1103/PhysRevC.92.061602}}.
\newline\urlprefix\url{https://link.aps.org/doi/10.1103/PhysRevC.92.061602}

\bibitem{Jin18}
J.~Lei, \href{https://link.aps.org/doi/10.1103/PhysRevC.97.034628}{Inclusive
  breakup calculations in angular momentum basis: Application to
  $^{7}\mathrm{Li}+^{58}\mathrm{Ni}$}, Phys. Rev. C 97 (2018) 034628.
\newblock \href {http://dx.doi.org/10.1103/PhysRevC.97.034628}
  {\path{doi:10.1103/PhysRevC.97.034628}}.
\newline\urlprefix\url{https://link.aps.org/doi/10.1103/PhysRevC.97.034628}

\bibitem{Jin18b}
J.~Lei, A.~M. Moro,
  \href{https://link.aps.org/doi/10.1103/PhysRevC.97.011601}{Post-prior
  equivalence for transfer reactions with complex potentials}, Phys. Rev. C 97
  (2018) 011601.
\newblock \href {http://dx.doi.org/10.1103/PhysRevC.97.011601}
  {\path{doi:10.1103/PhysRevC.97.011601}}.
\newline\urlprefix\url{https://link.aps.org/doi/10.1103/PhysRevC.97.011601}

\bibitem{Pot17}
G.~Potel, G.~Perdikakis, B.~V. Carlson, M.~C. Atkinson, W.~H. Dickhoff, J.~E.
  Escher, M.~S. Hussein, J.~Lei, W.~Li, A.~O. Macchiavelli, A.~M. Moro, F.~M.
  Nunes, S.~D. Pain, J.~Rotureau,
  \href{https://doi.org/10.1140/epja/i2017-12371-9}{Toward a complete theory
  for predicting inclusive deuteron breakup away from stability}, Eur. Phys. J.
  A 53~(9) (2017) 178.
\newblock \href {http://dx.doi.org/10.1140/epja/i2017-12371-9}
  {\path{doi:10.1140/epja/i2017-12371-9}}.
\newline\urlprefix\url{https://doi.org/10.1140/epja/i2017-12371-9}

\bibitem{Carlson2016}
B.~V. Carlson, R.~Capote, M.~Sin,
  \href{http://dx.doi.org/10.1007/s00601-016-1054-8}{Inclusive proton emission
  spectra from deuteron breakup reactions}, Few-Body Syst. 57~(5) (2016)
  307--314.
\newblock \href {http://dx.doi.org/10.1007/s00601-016-1054-8}
  {\path{doi:10.1007/s00601-016-1054-8}}.
\newline\urlprefix\url{http://dx.doi.org/10.1007/s00601-016-1054-8}

\bibitem{Pot15}
G.~Potel, F.~M. Nunes, I.~J. Thompson,
  \href{http://0-link.aps.org.fama.us.es/doi/10.1103/PhysRevC.92.034611}{Establishing
  a theory for deuteron-induced surrogate reactions}, Phys. Rev. C 92 (2015)
  034611.
\newblock \href {http://dx.doi.org/10.1103/PhysRevC.92.034611}
  {\path{doi:10.1103/PhysRevC.92.034611}}.
\newline\urlprefix\url{http://0-link.aps.org.fama.us.es/doi/10.1103/PhysRevC.92.034611}

\bibitem{bb}
A.~Bonaccorso, D.~M. Brink,
  \href{https://link.aps.org/doi/10.1103/PhysRevC.38.1776}{Nucleon transfer to
  continuum states}, Phys. Rev. C 38 (1988) 1776--1786.
\newblock \href {http://dx.doi.org/10.1103/PhysRevC.38.1776}
  {\path{doi:10.1103/PhysRevC.38.1776}}.
\newline\urlprefix\url{https://link.aps.org/doi/10.1103/PhysRevC.38.1776}

\bibitem{hasan}
H.~Hasan,
  \href{https://ora.ox.ac.uk/?utf8=%E2%9C%93&q=Hasan%2C+Hashima&search_field=author}{Study
  of reactions between heavy nuclei}, Ph.D. thesis, University of Oxford
  (1976).
\newline\urlprefix\url{https://ora.ox.ac.uk/?utf8=%E2%9C%93&q=Hasan%2C+Hashima&search_field=author}

\bibitem{Hasan_1979}
H.~Hasan, D.~M. Brink,
  \href{https://doi.org/10.1088%2F0305-4616%2F5%2F6%2F005}{The transfer
  amplitude and angular distributions in heavy-ion reactions}, Journal of
  Physics G: Nuclear Physics 5~(6) (1979) 771--779.
\newblock \href {http://dx.doi.org/10.1088/0305-4616/5/6/005}
  {\path{doi:10.1088/0305-4616/5/6/005}}.
\newline\urlprefix\url{https://doi.org/10.1088%2F0305-4616%2F5%2F6%2F005}

\bibitem{Monaco_1985}
L.~{Lo Monaco}, D.~M. Brink,
  \href{https://doi.org/10.1088%2F0305-4616%2F11%2F8%2F010}{Perturbation
  approach to nucleon transfer in heavy-ion reactions}, Journal of Physics G:
  Nuclear Physics 11~(8) (1985) 935--952.
\newblock \href {http://dx.doi.org/10.1088/0305-4616/11/8/010}
  {\path{doi:10.1088/0305-4616/11/8/010}}.
\newline\urlprefix\url{https://doi.org/10.1088%2F0305-4616%2F11%2F8%2F010}

\bibitem{27}
A.~Bonaccorso, D.~M. Brink, L.~L. Monaco,
  \href{https://doi.org/10.1088%2F0305-4616%2F13%2F11%2F013}{Nucleon transfer
  in heavy-ion reactions: energy dependence of the cross section}, Journal of
  Physics G: Nuclear Physics 13~(11) (1987) 1407--1428.
\newblock \href {http://dx.doi.org/10.1088/0305-4616/13/11/013}
  {\path{doi:10.1088/0305-4616/13/11/013}}.
\newline\urlprefix\url{https://doi.org/10.1088%2F0305-4616%2F13%2F11%2F013}

\bibitem{Winfield85}
J.~Winfield, N.~Jelley, W.~Rae, C.~Woods,
  \href{http://www.sciencedirect.com/science/article/pii/0375947485902271}{Spectroscopic-factor
  discrepancies in (9be, 10b) for different ejectile excitations}, Nuclear
  Physics A 437~(1) (1985) 65 -- 92.
\newblock \href
  {http://dx.doi.org/https://doi.org/10.1016/0375-9474(85)90227-1}
  {\path{doi:https://doi.org/10.1016/0375-9474(85)90227-1}}.
\newline\urlprefix\url{http://www.sciencedirect.com/science/article/pii/0375947485902271}

\bibitem{Winfield89}
J.~S. Winfield, E.~Adamides, S.~M. Austin, G.~M. Crawley, M.~F. Mohar, C.~A.
  Ogilvie, B.~Sherrill, M.~Torres, G.~Yoo, A.~Nadasen,
  \href{https://link.aps.org/doi/10.1103/PhysRevC.39.1395}{$^{12}\mathrm{induced}$
  single particle transfer reactions at e/a=50 mev}, Phys. Rev. C 39 (1989)
  1395--1401.
\newblock \href {http://dx.doi.org/10.1103/PhysRevC.39.1395}
  {\path{doi:10.1103/PhysRevC.39.1395}}.
\newline\urlprefix\url{https://link.aps.org/doi/10.1103/PhysRevC.39.1395}

\bibitem{Pieper78}
S.~C. Pieper, M.~H. Macfarlane, D.~H. Gloeckner, D.~G. Kovar, F.~D. Becchetti,
  B.~G. Harvey, D.~L. Hendrie, H.~Homeyer, J.~Mahoney, F.~P\"uhlhofer, W.~von
  Oertzen, M.~S. Zisman,
  \href{http://link.aps.org/doi/10.1103/PhysRevC.18.180}{Energy dependence of
  elastic scattering and one-nucleon transfer reactions induced by
  $^{16}\mathrm{O}$ on $^{208}\mathrm{Pb}$. i}, Phys. Rev. C 18 (1978)
  180--204.
\newblock \href {http://dx.doi.org/10.1103/PhysRevC.18.180}
  {\path{doi:10.1103/PhysRevC.18.180}}.
\newline\urlprefix\url{http://link.aps.org/doi/10.1103/PhysRevC.18.180}

\bibitem{Olmer78}
C.~Olmer, M.~Mermaz, M.~Buenerd, C.~K. Gelbke, D.~L. Hendrie, J.~Mahoney, D.~K.
  Scott, M.~H. Macfarlane, S.~C. Pieper,
  \href{http://link.aps.org/doi/10.1103/PhysRevC.18.205}{Energy dependence of
  elastic scattering and one-nucleon transfer reactions induced by
  $^{16}\mathrm{O}$ on $^{208}\mathrm{Pb}$. ii}, Phys. Rev. C 18 (1978)
  205--222.
\newblock \href {http://dx.doi.org/10.1103/PhysRevC.18.205}
  {\path{doi:10.1103/PhysRevC.18.205}}.
\newline\urlprefix\url{http://link.aps.org/doi/10.1103/PhysRevC.18.205}

\bibitem{PhysRevC.84.044616}
A.~M. Mukhamedzhanov,
  \href{https://link.aps.org/doi/10.1103/PhysRevC.84.044616}{Theory of deuteron
  stripping: From surface integrals to a generalized $r$-matrix approach},
  Phys. Rev. C 84 (2011) 044616.
\newblock \href {http://dx.doi.org/10.1103/PhysRevC.84.044616}
  {\path{doi:10.1103/PhysRevC.84.044616}}.
\newline\urlprefix\url{https://link.aps.org/doi/10.1103/PhysRevC.84.044616}

\bibitem{PhysRevC.89.054605}
J.~E. Escher, I.~J. Thompson, G.~Arbanas, C.~Elster, V.~Eremenko, L.~Hlophe,
  F.~M. Nunes,
  \href{https://link.aps.org/doi/10.1103/PhysRevC.89.054605}{Reexamining
  surface-integral formulations for one-nucleon transfers to bound and
  resonance states}, Phys. Rev. C 89 (2014) 054605.
\newblock \href {http://dx.doi.org/10.1103/PhysRevC.89.054605}
  {\path{doi:10.1103/PhysRevC.89.054605}}.
\newline\urlprefix\url{https://link.aps.org/doi/10.1103/PhysRevC.89.054605}

\bibitem{BAUR1984333}
G.~Baur, F.~Rösel, D.~Trautmann, R.~Shyam,
  \href{http://www.sciencedirect.com/science/article/pii/0370157384901388}{Fragmentation
  processes in nuclear reactions}, Physics Reports 111~(5) (1984) 333 -- 371.
\newblock \href
  {http://dx.doi.org/https://doi.org/10.1016/0370-1573(84)90138-8}
  {\path{doi:https://doi.org/10.1016/0370-1573(84)90138-8}}.
\newline\urlprefix\url{http://www.sciencedirect.com/science/article/pii/0370157384901388}

\bibitem{bb1}
A.~Bonaccorso, D.~M. Brink,
  \href{https://link.aps.org/doi/10.1103/PhysRevC.43.299}{Absorption versus
  breakup in heavy-ion reactions}, Phys. Rev. C 43 (1991) 299--310.
\newblock \href {http://dx.doi.org/10.1103/PhysRevC.43.299}
  {\path{doi:10.1103/PhysRevC.43.299}}.
\newline\urlprefix\url{https://link.aps.org/doi/10.1103/PhysRevC.43.299}

\bibitem{67}
A.~Bonaccorso, D.~M. Brink,
  \href{https://link.aps.org/doi/10.1103/PhysRevC.44.1559}{Stripping to the
  continuum of $^{208}\mathrm{Pb}$}, Phys. Rev. C 44 (1991) 1559--1568.
\newblock \href {http://dx.doi.org/10.1103/PhysRevC.44.1559}
  {\path{doi:10.1103/PhysRevC.44.1559}}.
\newline\urlprefix\url{https://link.aps.org/doi/10.1103/PhysRevC.44.1559}

\bibitem{initst}
A.~Bonaccorso,
  \href{https://link.aps.org/doi/10.1103/PhysRevC.60.054604}{Initial state
  dependence of the breakup of weakly bound carbon isotopes}, Phys. Rev. C 60
  (1999) 054604.
\newblock \href {http://dx.doi.org/10.1103/PhysRevC.60.054604}
  {\path{doi:10.1103/PhysRevC.60.054604}}.
\newline\urlprefix\url{https://link.aps.org/doi/10.1103/PhysRevC.60.054604}

\bibitem{ppnp}
A.~Bonaccorso,
  \href{http://www.sciencedirect.com/science/article/pii/S014664101830005X}{Direct
  reaction theories for exotic nuclei: An introduction via semi-classical
  methods}, Progress in Particle and Nuclear Physics 101 (2018) 1 -- 54.
\newblock \href {http://dx.doi.org/https://doi.org/10.1016/j.ppnp.2018.01.005}
  {\path{doi:https://doi.org/10.1016/j.ppnp.2018.01.005}}.
\newline\urlprefix\url{http://www.sciencedirect.com/science/article/pii/S014664101830005X}

\bibitem{Ravinder}
R.~Kumar, A.~Bonaccorso, Dynamical effects in proton breakup from exotic
  nuclei, Phys. Rev. C 84 (2011) 014613.
\newblock \href {http://dx.doi.org/10.1103/PhysRevC.84.014613}
  {\path{doi:10.1103/PhysRevC.84.014613}}.

\bibitem{firk}
F.~W.~K. Firk, Introduction to relativistic collisions\href
  {http://dx.doi.org/https://arxiv.org/abs/1011.1943}
  {\path{doi:https://arxiv.org/abs/1011.1943}}.

\bibitem{jer2}
J.~Margueron, A.~Bonaccorso, D.~Brink, A non-perturbative approach to halo
  breakup, Nuclear Physics A 720~(3) (2003) 337 -- 353.
\newblock \href
  {http://dx.doi.org/https://doi.org/10.1016/S0375-9474(03)01092-3}
  {\path{doi:https://doi.org/10.1016/S0375-9474(03)01092-3}}.

\end{thebibliography}
\end{document}

\title{Supplemental material for ``Comparison of semiclassical transfer to continuum model with Ichimura-Austern-Vincent model in medium energy knockout reactions''}
%Alt: Insights into the reaction mechanisms leading to partial fusion of weakly bound nuclei

% repeat the \author .. \affiliation  etc. as needed
% \email, \thanks, \homepage, \altaffiliation all apply to the current
% author. Explanatory text should go in the []'s, actual e-mail
% address or url should go in the {}'s for \email and \homepage.
% Please use the appropriate macro foreach each type of information

% \affiliation command applies to all authors since the last
% \affiliation command. The \affiliation command should follow the
% other information
% \affiliation can be followed by \email, \homepage, \thanks as well.
\author{Jin Lei  \fnref{jinfootnote} }
\author{Angela Bonaccorso\fnref{myfootnote}}
%\tnoteref{mytitlenote}}
%\tnotetext[mytitlenote]{jin.lei@pi.infn.it, bonac@df.unipi.it}
\fntext[jinfootnote]{{ jin.lei@pi.infn.it}}

\fntext[myfootnote]{{ bonac@df.unipi.it}}

%\homepage[]{Your web page}
%\thanks{}

%\altaffiliation{Present address: Institute of Nuclear and Particle Physics, and Department of Physics and Astronomy, Ohio University, Athens, Ohio 45701, USA}
%\homepage[]{Your web page}
%\thanks{}
\address{Istituto Nazionale di Fisica Nucleare, Sezione di Pisa, Largo Pontecorvo 3, 56127 Pisa, Italy.}

%\homepage[]{Your web page}
%\thanks{}

%\address{Istituto Nazionale di Fisica Nucleare, Sezione di Pisa, Largo Pontecorvo 3, 56127 Pisa, Italy}

%Collaboration name if desired (requires use of superscriptaddress
%option in \documentclass). \noaffiliation is required (may also be
%used with the \author command).
%\collaboration can be followed by \email, \homepage, \thanks as well.
%\collaboration{}
%\noaffiliation

% 25.70.Mn, Projectile and target fragmentation
% 24.10.Eq 	Coupled-channel and distorted-wave models
% 25.45.-z  2H-induced reactions
% 24.87.+y 	Surrogate reactions

% insert suggested PACS numbers in braces on next line
%Pacs: 24.10.Eq, 25.70.Mn, 25.45.-z
% insert suggested keywords - APS authors don't need to do this
%\keywords{}
\date{\today}%
%\maketitle must follow title, authors, abstract, \pacs, and \keywords
\maketitle

\section{Theoretical models}
\subsection{IAV model}
In this section, we briefly summarize the model of Ichimura, Austern, and Vincent (IAV), whose original derivation can be found in Refs.~\cite{IAV85,AUSTERN1987125} and has been also recently revisited by several authors~\cite{Jin15,Jin15b,Jin18,Jin18b,Pot17,Carlson2016,Pot15}. We outline here the main results of this model and refer the reader to these references for further details on their derivations.

We write the process under study in the form,
\begin{equation}
P(=C+n)+T \rightarrow C+B^{*}, 
\end{equation}

where the projectile $P$, composed of $C$ and $n$, collides
with a target $T$, emitting $C$ fragments and any other fragments. Thus, $B^{*}$ denotes any final state of the $n+T$ system.

\begin{figure}[tb]
\begin{center}
 {\centering \resizebox*{0.56\columnwidth}{!}{\includegraphics{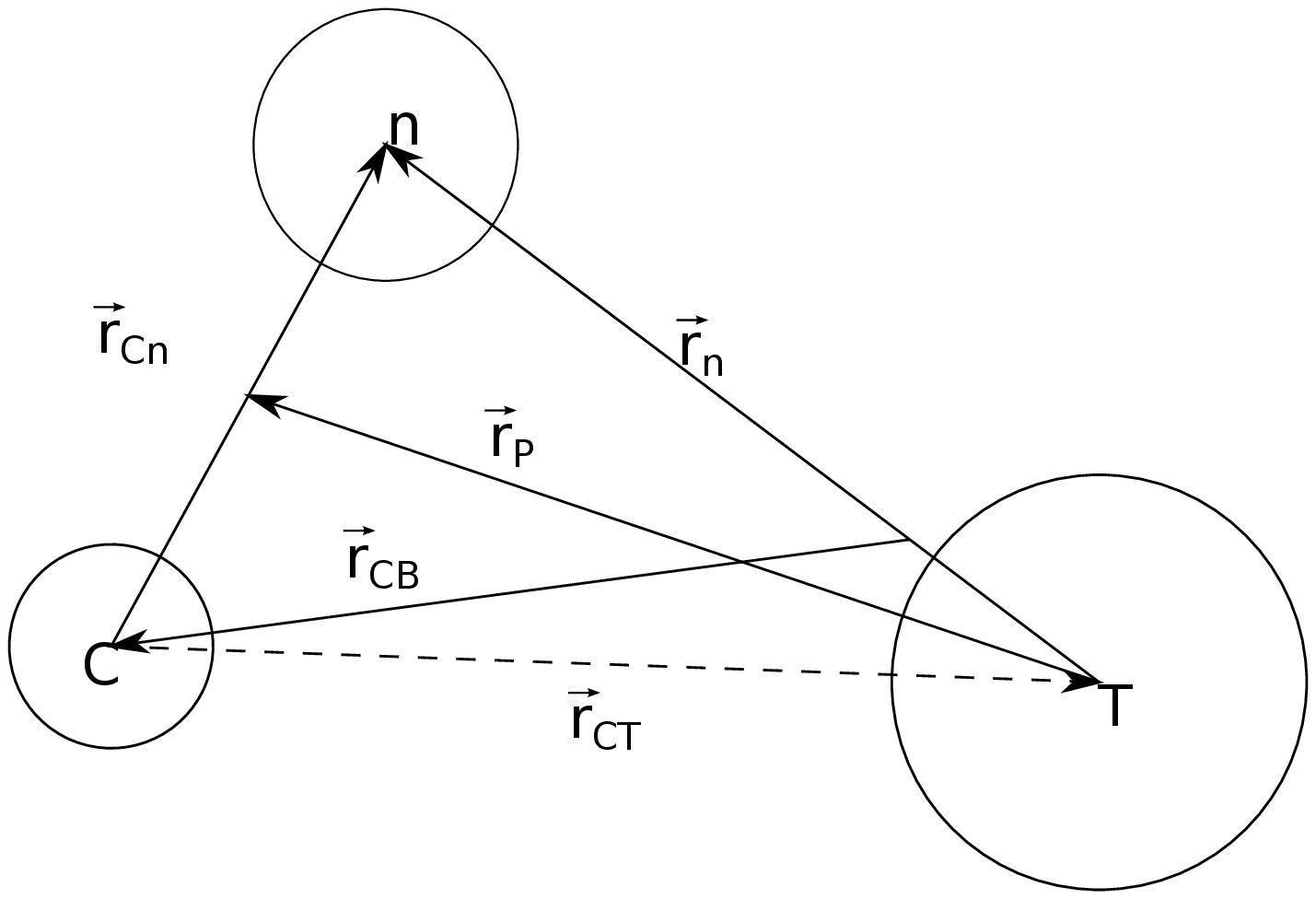}} \par}
\caption{\label{fig:IAV_coordinates}Coordinates used in the IAV model.}
\end{center}
\end{figure}

This process will be described with the effective Hamiltonian
\begin{equation}
\label{eq:IAV_H}
H=K+V_{C n}+U_{C T}\left(\mathbf{r}_{CT}\right)+H_{T}(\xi)+V_{n T}\left(\xi, \mathbf{r}_{n}\right),
\end{equation}
where $K$ is the total kinetic energy operator, $V_{Cn}$ is 
the interaction binding the two clusters $C$ and $n$ in the 
initial composite nucleus $P$, $H_T(\xi)$ is the Hamiltonian 
of the target nucleus (with $\xi$ denoting its internal 
coordinates), and $V_{nT}$ and $U_{CT}$ are the fragment–target interactions. The relevant coordinates are depicted in Fig.~\ref{fig:IAV_coordinates}.

In writing the Hamiltonian of the system in the form~(\ref{eq:IAV_H}) we make a clear distinction between the 
two cluster constituents; the interaction of the fragment 
$C$, the one that is assumed to be observed in the 
experiment, with the target, is described with a (complex) optical potential.
Inclusive breakup processes arising from this interaction
(e.g., target excitation) are included only effectively
through $U_{CT}$. The particle $C$ is said to act as 
spectator. On the other hand, the interaction of the particle $n$ with the target retains the dependence of the target degrees of freedom ($\xi$). In the following this dependence is released by the choice of a neutron-target optical potential.

Starting from the  Hamiltonian of Eq.~(\ref{eq:IAV_H}) IAV derived the 
following expression for the double differential cross 
section for the inclusive breakup with respect to the angle 
and energy of the $C$ fragments:

% \begin{align}
% \label{eq:iav}
% \frac{d^2\sigma}{dE_C d\Omega_C} =  \frac{2\pi}{\hbar v_p}\rho_C(E_C)\Big( \int |\langle \chi_n^{(-)} | \rho_n  \rangle|^2 \delta(E^{3b}-E_C-E_n) d\mathbf{k}_n 
% - \langle G_n\rho_n | W_n| G_n \rho_n \rangle /\pi \Big)
% \end{align}

\begin{align}
\label{eq:iav}
\frac{d^2\sigma}{dE_C d\Omega_C} =  \frac{2\pi}{\hbar v_P}\rho_C(E_C)\Big( \rho_n(E_n)\int |\langle \chi_n^{(-)} | \mathcal{S}_n  \rangle|^2 d\Omega_n 
- \langle G_n\mathcal{S}_n | W_n| G_n \mathcal{S}_n \rangle /\pi \Big)
\end{align}

where $v_P$ is the projectile-target relative velocity, $\rho_C(E_b) =
k_C\mu_C/((2\pi)^3\hbar^2)$ and $\rho_n(E_n) =
k_n\mu_n/((2\pi)^3\hbar^2)$ are the density of states for the particle $C$ and $n$ respectively, $W_n$ is the imaginary part of the optical potential $U_n$, obtained from the optical reduction of $V_{n T}\left(\xi, \mathbf{r}_{n}\right)$ which 
describes $n + T$ elastic scattering, $G_n= {1}/({E_n^{(+)}-U_n-T_n })$
is the Green's function for the neutron-target channel, $E_n$ is the energy in the $n+T$ channel which satisfies the energy conservation, $\chi_n$ is the distorted-wave in this channel,
and $\mathcal{S}_n$ is the source term which takes the form 
\begin{equation}
\label{eq:source}
\mathcal{S}_n(\mathbf{r}_n) = \langle \mathbf{r}_n\chi_C^{(-)} |V_{post} |\Psi^{3b(+)} \rangle ,
\end{equation}
$\chi_C^{(-)}$is the distorted-wave describing the scattering of $C$ in the final channel with respect to the $n + T$ subsystem, and $V_{post} = V_{Cn} + U_{CT} - U_C$ (with $U_C$ the
optical potential in the final channel) is the post form transition
operator.
One should note that there is a natural separation between the elastic part and nonelastic part in Eq.~(\ref{eq:iav}), the first part corresponds to the elastic interaction between the $n$ and $T$ which is called elastic breakup (EBU), whereas the second term accounts for the cross section of the nonelastic process named nonelastic breakup (NEB).

Austern et al.~\cite{AUSTERN1987125} suggested approximating the three-body
wave function appearing in the source term of Eq.~(\ref{eq:source}), 
$\Psi^{3b(+)}$, by the CDCC one. Since the CDCC wave function is also
a complicated object by itself, a simpler choice is to use
the distorted-wave Born approximation (DWBA), i.e., $\Psi^{3b(+)}= \chi^{(+)}_{PT}(\mathbf{r}_P)\phi_P(\mathbf{r}_{Cn})$, where $\chi^{(+)}_{PT}(\mathbf{r}_P)$ is a distorted wave describing
$P + T$ elastic scattering and $\phi_P(\mathbf{r}_{Cn})$ is the projectile ground-state wave function.

\subsection{TC model}

The semiclassical TC model \cite{bb} is a generalization to final unbound states of the transfer between bound states model of Brink and collaborators \cite{hasan,Hasan_1979,Monaco_1985,27}. Semiclassical methods were very popular in the '70s as a substitute to full DWBA calculations which, in those days, were very lengthy  and computationally expensive. Transfer reactions were a common tool to study single particle  characteristics, in particular occupation probabilities but often  theoretical calculations  gave cross sections much larger than the data and spectroscopic factors different from the shell model  values~\cite{Winfield85,Winfield89,Pieper78,Olmer78}. Then an attempt to disentangle the content and the ingredients of the DWBA approach via semiclassical methods which could provide analytical expressions for the cross sections. The procedure followed was first to choose a  WKB wave functions for the distorted waves, then the standard reduction of the three dimensional integral for the transfer form factor to a surface integral, similarly to what is done in Refs.~\cite{PhysRevC.84.044616,PhysRevC.89.054605,BAUR1984333}. Finally analytical Hankel functions were used for the initial and final states on the surface between the two nuclei. The method is valid for peripheral reactions based on the core spectator model as mentioned above.

One of the advantages of a semiclassical method like the TC is that it allows a transparent interpretation of the formalism and of its results, therefore we remind in the following a few steps leading to the final probability and cross section formulae Eqs.~(\ref{dpde}) and~(\ref{totx}). Apart from the original references \cite{bb,bb1,67,initst} a detailed account of the final formulae derivation can be found in the review paper \cite{ppnp}.

 The semiclassical transfer to the continuum amplitude \cite{bb} calculated in the target reference frame is: 
\begin{equation}
A_{fi}=\frac{1}{ i\hbar}
\int_{-\infty}^{\infty}dt<\phi_{f} ({\bf r_n})|U_{nT}({\bf r_n})|\phi_{i}({\bf r_n-R}(t))>e^{-i(\omega t-mvz/\hbar)},\label{1}
\end{equation}
Where the time dependent nucleon initial and final wave functions in their respective reference frames are  $\psi_i({\bf r_n^{\prime}},t)=\phi_{i} ({\bf r_n^{\prime}})e^{-i\varepsilon_i t/ \hbar}$ and $\psi_f({\bf r_n},t)=\phi_{f} ({\bf r_n})e^{-i\varepsilon_f t/ \hbar}$.
 In Eq.~(\ref{1}),  then the initial-state wave  function has to be boosted by a Galilean transformation. This implies  a change of variables by the introduction of  ${\bf R}(t)={\bf b_c}+vt$, the classical trajectory of relative motion between projectile and target. We work in a reference frame in which ${\bf r_P} ={\bf r_{CT}}= {\bf R}(t)$ in Fig.1. $U_{nT}({\bf r_n})$ is the nucleon target interaction in the final state, corresponding to $V_{nT}({\xi , \bf r_n})$ of Eq.(\ref{eq:IAV_H}) after the optical reduction.
Initial and final radial wave functions are taken as Hankel functions according to \cite{bb}. 
For proton breakup we use the same type of wave functions. For the initial state we first calculate the exact proton  bound state wave function and then we fit to it a neutron wave function. Finally  we use such a neutron wave function and the corresponding, effective separation energy.  This method was checked and found very accurate in Ref.\cite{Ravinder}.

With the the above choices and after some analytical manipulation the modulus square of  the amplitude Eq.(\ref{1})  leads to the breakup probability:

\begin{equation} 
\frac{dP_{-n} }{d\zeta}\approx \frac{1}{2}\Sigma_{j_f}(|1-\bar S_{j_f} |^2+1-|\bar S_{j_f} |^2)
(2j_f+1)(1+B_{if})
\left [\frac{\hbar}{mv}\right ]
\frac{1}{f}
|C_i|^2 
\frac{e^{-2\eta b_c}}{2\eta b_c}
M_{l_fl_i}, \label{dpde}
\end{equation} 
where $\zeta$ can be the final nucleon  energy with respect to the target ($\varepsilon_f$) or the  nucleon momentum in the initial (k$_1$) or final (k$_2$) state.
  $\bar S_{j_f}$ are neutron-target S-matrices calculated for each neutron final energy according to the optical model in an energy dependent optical potential, including the spin-orbit term of the neutron-target optical potential. The sum of the two terms $(|1-\bar S_{j_f} |^2+1-|\bar S_{j_f} |^2)$ is obtained automatically \cite{bb} as a result of using an unitary  energy averaged optical model S-matrix in the definition of the final continuum state  thus  including  non elastic and elastic breakup (stripping and diffraction). These two terms correspond to the first and second term   of  Eq.(\ref{eq:iav}). The sum over partial waves in Eq.(\ref{dpde}) is  a sum over total   neutron-target angular momenta. $C_i$ is the initial wave-function asymptotic normalization constant. It is obtained as the ratio between the  numerically calculated single particle wave function and the Hankel function. The form factor  $\frac{e^{-2\eta b_c}}{2\eta b_c}$is due to the combined effects of the initial and final wave-function Fourier transforms,  while $M_{l_fl_i}$ is due to the overlap of the angular parts. B$_{if}$ are spin-coupling coefficients. Further definitions and discussion can be found in Refs.\cite{initst,ppnp}.

By using 4-energy momentum conservation (see for example \cite{firk}) and the relative Jacobian,  the differential $d{\sigma_{TC}/ {d\zeta}}$ cross section becomes directly comparable to the measured momentum distributions function of P$_{//}$ the core parallel momentum. 
Finally the total breakup  cross section is obtained  by integrating over energy or momentum the differential breakup probability and then on the core-target impact parameter by  weighting it with the probability $|S_{ct}(b_c)|^2$ that the  measured core
has survived  "intact" the scattering, and by multiplying it by the spectroscopic factor of the initial state $C^2S$ if a shell model Woods-Saxon wave functions used:

\begin{equation}\sigma _{TC}=C^2S
   \int_0^{\infty} d{\bf b_c}   |S_{ct}(b_c)|^2\int {\frac {d P_{-n}(b_c)} {d\zeta} } d\zeta  .\label{totx}\end{equation}

 This formalism does not include Coulomb recoil effects of the core because it does not distinguish the center of mass of the core-target system from the center of mass of the projectile-target. Core recoil effects give rise to the so called Coulomb breakup which is important for heavy targets \cite{jer2,ppnp}.
As we shall see in the following section the IAV and TC methods lead to very close results. However from the formal point of view they look quite different if one compares Eq.(\ref{eq:iav}) with Eqs.(\ref{dpde})
and (\ref{totx}).  The main differences are that in the IAV method the core-target $S$-matrix is included in the source term Eq.(\ref{eq:source}) and and n-target $S$-matrix is considered as solution of an  inhomogeneous equation while in TC they are solutions of homogeneous equations. In practice this means that the TC method decouples the core-target scattering from the neutron-target scattering, considering them as independent. This  corresponds to consider off-shell effects negligible and thus it calculates  on-shell S-matrices. The historical origin of this difference lies in the fact that IAV approach originated as a method to calculate light nuclei ($d$) breakup at low energy while TC was developed to treat heavy-ion reactions at intermediate to high energies where surface approximation and thus decoupling of core-valence-particle degrees of freedom
appear as the natural choice.

\bibliography{references}